\documentclass[preprintnumbers,superscriptaddress,amsmath,amssymb,prd,preprint]{revtex4-1}

\usepackage[colorlinks,linkcolor=red,anchorcolor=blue,citecolor=green]{hyperref}

\usepackage{amsmath}
\usepackage{mathtools}
\usepackage{amssymb}
\usepackage{mathrsfs}
\usepackage{amsthm}
\usepackage{amsfonts}
\usepackage{cases}
\usepackage{bm}
\usepackage{bbm}
\usepackage{physics}

\usepackage{dcolumn}
\usepackage{enumerate}

\usepackage{graphics}
\usepackage{graphicx}
\usepackage{graphicx}
\usepackage{picinpar}
\usepackage{subfig}

\begin{document}	

\title{Accretion of a Plasma Vlasov Gas onto a Reissner--Nordstr\"om Black Hole}

\author{Yongqiang Liu}
\thanks{Electronic address: \href{mailto:liuyq8128@outlook.com}{liuyq8128@outlook.com}}
\affiliation{School of Physics and Technology, University of Jinan,
336 West Road of Nan Xinzhuang, Jinan, Shandong 250022, China}

\author{Hongsheng Zhang}
\thanks{Electronic address: \href{mailto:sps_zhanghs@ujn.edu.cn}{sps\_zhanghs@ujn.edu.cn (corresponding author)}}
\affiliation{School of Physics and Technology, University of Jinan,
336 West Road of Nan Xinzhuang, Jinan, Shandong 250022, China}

\date{\today}

\begin{abstract}

We develop a steady-state, spherically symmetric accretion model for a two-component plasma Vlasov gas in the background of a Reissner--Nordstr\"om black hole. Analytical expressions in parametric form for the critical angular momentum and impact parameter are derived for each species, leading to a rigorous partition of the absorption and scattering domains in phase space.
For both monoenergetic and J\"uttner distributions, electromagnetic attraction enhances, while repulsion suppresses, the particle number density, energy density, pressure and accretion rates of individual charged components relative to the neutral case. After proper normalization, the total physical quantities of the plasma (under non-extreme conditions) can be expressed exactly, or to high accuracy, as weighted averages of the two species.
When the two components share identical distribution functions, accretion drives the Reissner--Nordstr\"om black hole naturally toward the Schwarzschild limit---a charge evolution that is absent in neutral-gas models. Numerical results further reveal that near the horizon the radial pressure is significantly smaller than the tangential pressure, sharply distinguishing the Vlasov gas from a perfect fluid. 
\end{abstract}

\maketitle

\section{Introduction}

The traditional study of accretion theory has long been based on the fluid approximation, where Bondi's pioneering work established the basic framework for spherically symmetric steady accretion~\cite{Bondi(1944),Bondi(1952)} and was extended to compact objects by Michel~\cite{Michel(1972)}. However, when the gas becomes tenuous and the collision frequency is extremely low, the fluid approximation is no longer applicable, and one must turn to a kinetic description based on the Vlasov equation~\cite{Rioseco(2017a)}. This kinetic model is particularly crucial for characterizing low-density, high-temperature plasma environments, which are widely present in the accretion flows around supermassive black holes at galactic centers; the relevant extreme physical conditions have been explored theoretically~\cite{Falcke(2013)} and subsequently probed by observations from the Event Horizon Telescope~\cite{EHT(2019a),EHT(2019b),EHT(2022a),EHT(2022b)}.

In recent years, a complete mathematical framework for the theory of Vlasov gases in curved spacetime has been constructed based on the tangent and cotangent bundles~\cite{Sarbach(2014a),Sarbach(2014b),Acuna(2022),Rioseco(2024)}, which forms the foundation for the studies below.

In the simplified model of neutral test particles (neglecting self-gravity), Rioseco and Sarbach first solved exactly and discussed the accretion of a Vlasov gas in Schwarzschild spacetime~\cite{Rioseco(2017a),Rioseco(2017b)}, and their results show that near the event horizon the radial pressure is much smaller than the tangential pressure. This anisotropy leads to an accretion rate of the Vlasov gas far below the prediction of the Bondi--Michel fluid model, providing a natural physical explanation for low-luminosity accretion flows. For a charged black hole, Cie{\'s}lik and Mach found that the charge of a Reissner--Nordstr\"om black hole reduces the accretion rate~\cite{Cieslik(2020)}.

Other typical accretion scenarios include accretion of a Vlasov gas onto a moving Schwarzschild black hole~\cite{Mach(2021),Mach2021a,Mach2022}; gas distributions with finite radius~\cite{Gamboa(2021)}; accretion in modified gravity theories~\cite{Liao(2022),Cai(2023),Zhang2025}; accretion onto generic static and spherically symmetric black holes, generalizing the standard Schwarzschild case~\cite{Momennia2025}; and differences in accretion rates caused by different distribution functions~\cite{Li(2026)}. Monte Carlo numerical methods for studying Vlasov gas accretion have also been developed~\cite{Mach(2023),Cieslik(2024)}.  Stationary axisymmetric Vlasov gas disks surrounding Schwarzschild black holes have been studied by Gabarrete and Sarbach~\cite{Gabarrete2023}.

In the study of axisymmetric black holes, Cie{\'s}lik et al. first discussed the accretion of gas confined to the equatorial plane of Kerr spacetime~\cite{Cieslik(2022)}; subsequently Li et al. studied the accretion per unit solid angle in the full Kerr spacetime~\cite{Li(2023)}. For Kerr black holes, Mach et al. systematically investigated the accretion of a neutral Vlasov gas onto a Kerr black hole in two complementary works: a comprehensive treatment of the full Kerr accretion problem~\cite{Mach(2026a)}, and a Bondi-type analysis in the kinetic regime~\cite{Mach(2026b)}. These works found that the spin parameter suppresses the mass and energy accretion rates, while accretion reduces the absolute value of the spin parameter. Parallel to these developments, Khan and Mach extended a Monte Carlo scheme for stationary solutions of the Vlasov equation to the Kerr spacetime, focusing on razor-thin kinetic disks in the equatorial plane~\cite{Khan(2026c)}, and constructed finite collisionless accretion disks~\cite{Khan2025}.

For charged gases, the test-particle model provides a useful starting point. Schroven et al.~\cite{Schroven2017} constructed an analytic model for charged dust accreting onto a Kerr--Newman black hole and showed that even a vanishingly small black hole charge can significantly affect the flow when $eQ\sim 1$, with the accretion disk edges depending strongly on the plasma charge. This charged model generalizes earlier neutral dust models for Schwarzschild~\cite{Mendoza2009,Tejeda2012} and Kerr~\cite{Tejeda2013} spacetimes. These dust models, however, treat the gas as noninteracting test particles rather than as a Vlasov gas. 
Li et al.~\cite{Li(2025)} studied the accretion of a charged Vlasov gas with a Fermi--Dirac-type distribution onto a Reissner--Nordstr\"om black hole and discussed its implications for the cosmic censorship conjecture. Subsequently, Liu~\cite{Liu(2026)} extended the analysis to a two-component plasma Vlasov gas in the Kerr--Newman spacetime under weak electromagnetic coupling, showing that the principal values of the accretion rates reduce to those of the neutral gas at leading order, while a residual charge current slowly neutralizes the black hole charge. On the numerical side, fully kinetic simulations of a finite-angular-momentum plasma accreting onto a  spinning black hole have also been performed by Mehlhaff et al.~\cite{Mehlhaff2026}.

As for fully self-consistent descriptions, the local well-posedness of the Maxwell--Vlasov system on curved spacetimes was established by Noutchegueme et al.~\cite{Noutchegueme2001,Noutchegueme2002}. Fajman and Seer constructed static spherically symmetric solutions of the Einstein--Vlasov--Maxwell system describing a Reissner--Nordstr\"om black hole surrounded by a charged Vlasov matter shell of finite support, with vacuum regions inside and outside the shell, so that the entire spacetime is fully self-consistent~\cite{Fajman2026}. Further studies of self-gravitating Vlasov systems can be found in~\cite{Rein1994,Andreasson2014,Ames2016,Andreasson2021,Jabiri2021,Jabiri2022} and the related reviews~\cite{Rein2023,Ames2024}. In this paper we adopt the test-field approximation, neglecting the gas self-field and self-gravity backreaction; for a dilute accreting gas, this is a standard and self-consistent approximation.

Building on these developments, we study the accretion of a two-component plasma Vlasov gas onto a Reissner--Nordstr\"om black hole with general electromagnetic coupling.
We provide a complete analytical characterization of the absorption and scattering domains for charged particles in a background electromagnetic field, with closed-form boundary functions essential for accurate phase-space integration. We derive analytic expressions for the particle current and stress-energy tensor for both a single-component charged gas and a two-component plasma, which reduce consistently to the neutral-gas results in the asymptotic and weak-coupling limits~\cite{Cieslik(2020)}, and obtain explicit formulas for the particle number and energy accretion rates that clarify the opposing roles of electromagnetic attraction and repulsion. Beyond extending spherical Vlasov accretion studies to charged plasmas, the analysis reveals a weighted-average structure that simplifies the description of two-component collisionless plasmas after proper normalization, offering a systematic simplification framework for multicomponent systems and serving as the leading-order approximation to a fully self-consistent Vlasov--Maxwell theory.

The paper is organized as follows. Sec.~\ref{Sec2} analyzes the dynamics of charged test particles in Reissner--Nordstr\"om spacetime and provides an analytical delineation of the absorption and scattering domains. Sec.~\ref{Sec3} reviews the Vlasov gas theory and computes the integral expressions for the particle current density and the stress-energy tensor. Sec.~\ref{Sec4} constitutes the core of our work: we specialize to a two-component plasma, compute the explicit integral forms of the number density, energy density, principal pressures, and accretion rates, and discuss the influence of the electromagnetic coupling on these quantities. Sec.~\ref{Sec5} systematically establishes the mathematical structure relating the whole plasma to its constituents. Sec.~\ref{Sec6} presents numerical results that confirm our analytical findings, including radial profiles of the densities and pressures, as well as the dependence of accretion rates on the black hole charge and the plasma parameters. Finally, Sec.~\ref{Sec7} summarizes our main conclusions and discusses possible extensions.

We use geometric units, i.e., the gravitational constant and the speed of light are set to $G=c=1$, and the Lorentzian signature is $[-1,+1,+1,+1]$.

\section{Kinematic analysis of the gas particles}\label{Sec2}

Consider a Reissner--Nordstr\"om (RN) black hole $(M, Q, g)$ with the line element in spherical coordinates
\begin{align}
\dd{s^{2}}=-h(r)\dd{t^{2}}+\frac{1}{h(r)}\dd{r^{2}}+r^{2}\pqty{\dd{\theta^{2}}+\sin^{2}\theta \dd{\varphi^{2}}},  \qquad h(r)\equiv 1-\frac{2M}{r}+\frac{Q^{2}}{r^{2}},  \label{metric}
\end{align}
where the outer horizon radius is $r_{H}=M+\sqrt{M^2-Q^2}$. The electromagnetic potential 1-form and field strength 2-form are, respectively,
\begin{align}
A &=-\frac{Q}{r}\dd{t}, & F&=\dd{A}=-\frac{Q}{r^{2}}\dd{t}\wedge \dd{r}.
\end{align}

The equation of motion for a charged test particle (mass $m$, charge $q$) reads
\begin{align}
p^{c}\nabla_{c}p^{a} = q F^{a}{}_{b} p^{b}, \label{equationmq}
\end{align}
where $p^{a} =\dv{x^{\mu}}{\lambda} \pqty{\pdv{x^{\mu}}}^{a}$ is the physical momentum of the particle, $\lambda=\tau/m$ ($\tau$ is the affine parameter), and $F^{a}{}_{b}\equiv g^{ac}F_{cb}$. The physical momentum 1-form is $p_{a} \coloneqq g_{ab} p^{a}$, and the canonical momentum 1-form is $\pi_{a} \coloneqq p_{a} + q A_{a}$. The two Killing vector fields of the background spacetime, $\partial_{t}$ and $\partial_{\varphi}$, yield the conserved canonical energy and canonical axial angular momentum, respectively:
\begin{align}
E &\coloneqq -\pi_{t} = -p_{t} +\frac{qQ}{r}, \label{KillingE}\\
L_{z} &\coloneqq \pi_{\varphi} = p_{\varphi}.\label{KillingLz}
\end{align}
In addition, the Hamiltonian and the square of the total orbital angular momentum are conserved along the worldline:
\begin{align}
H &\coloneqq \frac{1}{2} g^{ab}(\pi_{a}-qA_{a})(\pi_{b}-qA_{b})= \frac{1}{2} g^{ab}p_{a}p_{b} = -\frac{1}{2} m^{2}, \label{KillingH}\\
L^{2} &\coloneqq  p_{\theta}^{2} + \frac{p_{\varphi}^{2}}{\sin^{2}\theta}. \label{KillingL}
\end{align}
Outside the event horizon, Eq.~\eqref{KillingH} guarantees that the physical momentum $p^{a}$ is a timelike vector, and its future-directedness requires
\begin{align}
p^{t}&=h(r)^{-1}	\pqty{E-\frac{qQ}{r}} >0.
\end{align}

To simplify the subsequent calculations, we introduce a set of dimensionless parameters (without introducing new symbols):
\begin{align}
r &\leftarrow \frac{r}{M},& Q & \leftarrow \frac{Q}{M}, &   E &\leftarrow \frac{E}{m}, &  L &\leftarrow \frac{L}{mM}.
\end{align}
In the following text, when $q$ appears alone it represents the particle charge, while the product $qQ$ stands for the dimensionless combination $\frac{qQ}{mM}$. Accordingly, we define the electromagnetic coupling parameter
\begin{align}
k\coloneqq \frac{qQ}{mM}=qQ,  \label{kqQ}
\end{align}
where the physical meaning of the absolute value $|k|$ is the ratio of the electrostatic force to the gravitational force at infinity. Electromagnetic attraction, charge neutrality (no electromagnetic interaction), and electromagnetic repulsion correspond to $k<0$, $k=0$, and $k>0$, respectively.

The radial equation of motion can be obtained from Eq.~\eqref{KillingH} and the expressions for $E, L_{z}, L^{2}$:
\begin{align}
\pqty{ p^{r}}^{2}&=\pqty{E-k/r}^{2}-h(r) \pqty{ L^{2}/r^{2}+1} \equiv R(r). \label{dotr}
\end{align}
The complete geodesic solutions in terms of Weierstrass functions are given in Ref.~\cite{Grunau2011}; in what follows, we only consider orbits that can extend to spatial infinity. We define the effective energy $E_{\text{eff}}$ and the effective potential $U_{\text{eff}}$:
\begin{align}
E_{\text{eff}}(r, E)& \coloneqq E-k/r, &	U_{\text{eff}}(r, L) &\coloneqq \sqrt{h(r)}\sqrt{1+L^{2}/r^{2}}.
\end{align}
In the limit $r \to \infty$,
\begin{align}
E_{\text{eff}}-U_{\text{eff}}&\simeq (E-1)+\frac{(1-k)}{r}-\frac{L^{2}+Q^{2}-1}{2r^{2}},
\end{align}
which shows that when $k\geq 1$, a particle at rest ($E=1$) cannot fall from infinity to a finite radius. Since the subsequent discussion involves particles initially at rest at infinity, we restrict ourselves to the case $k<1$.

For a given energy $E \geq 1$, the set of angular momenta $L\equiv \sqrt{L^{2}}$ for which a particle is accreted (captured) by the black hole is
\begin{align}
D^{\text{abs}}_{L}(E)\coloneqq \Bqty{L \mid  \text{there is no solution of } E_{\text{eff}}<U_{\text{eff}} \text{ on } r\in (r_{H},+\infty)},
\end{align}
where obviously $\inf(D^{\text{abs}}_{L})=0$; we define $L_{c}(E,k)\coloneqq \sup(D^{\text{abs}}_{L})$. When $L>L_{c}(E, k)$, the orbit possesses a radial turning point $r_\text{turn}$. Hence, the set of angular momenta for all scattering orbits passing through a spacetime point $x\in M$ is
\begin{align}
D^{\text{scat}}_{L}(E, x)\coloneqq \Bqty{L > L_{c}(E,k) \mid \text{there is no solution of } E_{\text{eff}}<U_{\text{eff}} \text{ on } r\in (r_{x},+\infty)},
\end{align}
with $L_{c}(E,k) =\inf(D^{\text{scat}}_{L})$. We denote $ L_{\max}\coloneqq\sup(D^{\text{scat}}_{L}(E, x))$, which is the maximum allowed angular momentum at $r_{x}$:
\begin{align}
L_{\max}(r, E)= \sqrt{h(r)^{-1}\pqty{Er-k}^{2}-r^{2}}.
\end{align}

For the critical angular momentum $L_{c}(E,k)$, $E_{\text{eff}}$ and $U_{\text{eff}}$ are tangent at a certain position $r_{c}$; solving the corresponding equations yields
\begin{align}
E_{\text{solve}}(r, k)&=\frac{k}{r} +\frac{1}{r}\sqrt{\frac{h}{B^{2}}}, \\
L_{\text{solve}}(r, k)&=\sqrt{\frac{1}{B^{2}}-r^{2}},\\
B_{\pm}&=\frac{ k\pm \sqrt{k^{2}+4 P(r)}}{2r^{2} \sqrt{h}},
\end{align}
where $P(r)\equiv r^{2}-3r+2Q^{2}$. The branch selection for $B$ is as follows:
\begin{align}
B(r, k)=\begin{cases}
B^{(1)}_{+}, &r\in (r_{\text{ph}}, r_{\text{mb}}(k)], 	 \quad k \leq 0, \\
B^{(2)}_{+}, &r\in [r_{\min}(k), r_{\text{mb}}(k)], 	 \quad k \in (0, 1),\\
B^{(2)}_{-}, &r\in [r_{\min}(k), r_{\text{ph}}), 	\quad k \in (0,1).
\end{cases}
\end{align}
For electromagnetic attraction or charge neutrality ($k \leq 0$), only the $B^{(1)}_{+}$ branch exists, corresponding to $E\in [1, +\infty)$; for electromagnetic repulsion ($k\in (0,1)$), the $B^{(2)}_{+}$ branch corresponds to the low-energy branch $E\in [1, E_{0}]$, and $B^{(2)}_{-}$ corresponds to the high-energy branch $E\in [E_{0}, +\infty)$. The boundaries are defined as
\begin{align}
r_{\text{mb}}(k)& = \frac{4}{3}+\frac{2}{3}\sqrt{\frac{\Delta(k)}{1-k}}\cos\bqty{\frac{1}{3}\arccos\pqty{\frac{D(k)\sqrt{1-k}}{2\Delta^{3/2}}}}, \quad k<1,\\
r_{\text{ph}}&\equiv \frac{3}{2}\pqty{1+\sqrt{1-\frac{8Q^{2}}{9}}},\label{rphQ}\\
r_{\min}(k)&=\frac{3}{2}\pqty{1+\sqrt{1-\frac{8Q^{2}+k^{2}}{9}}}, \quad k \in (0,1), \label{rminQk}\\
 E_{0}&\equiv E_{\text{solve}}(r_{\min})=\frac{k}{r_{\min}}+\frac{2r_{\min}}{k}h(r_{\min}),
\end{align}
where
\begin{align*}
\Delta(k) &\equiv 4(4-3Q^{2})-k(16-9Q^{2})+3k^{2},\\
D(k)&\equiv 27Q^{4}-144Q^{2}+128+k(108Q^{2}-128)+k^{2}(36-27Q^{2}).
\end{align*}
The physical meaning of the above quantities (all $r$ outside the event horizon) is as follows: $r_{\text{ph}}$ is the radius of the photon sphere ($P=0$); $r_{\text{mb}}(k)$ is the root of $E_{\text{solve}}(r, k)=1$; for electromagnetic repulsion, $r_{\min}(k)$ is the root of $k^{2}+4P=0$, which ensures the continuity of the two branches of $E_{\text{solve}}(r, k)$, and the energy at the connection point of the two branches is $E_{0}$.

The critical scattering condition (solution of a quadratic equation) guarantees a one-to-one correspondence between $(E_{\text{solve}}, L_{\text{solve}}, r)$ on each branch of $B(r,k)$. What we are truly interested in is the critical scattering position $r_{c}(E, k)$ for a given energy $E$, and the critical curve in phase space
\begin{align}
L_{c}(E, k)\coloneqq L_{\text{solve}}(r_{c}(E, k), k), \quad E\in [1, +\infty), \quad k<1.
\end{align}
For electromagnetic attraction and charge neutrality, $r_{c}(E, k)$ and $L_{c}(E, k)$ are both $C^{1}$ functions of $E$. We now show that they are also of class $C^{1}$ in the electromagnetic repulsion case. In fact, it suffices to prove that $r_{c}(E, k)$ is differentiable at $E_{0}$.
\begin{align}
\mathop{\mathrm{lim}}_{r \to r^{+}_{\min}}\frac{\abs{E(r, k)-E_{0}}}{r-r_{\min}}=\mathop{\mathrm{lim}}_{r \to r^{+}_{\min}}\frac{\sqrt{k^{2}+4P}}{2(r^{2}-3r+2Q^{2})(r-r_{\min})}\to +\infty,
\end{align}
in the last step, unless $k=\sqrt{9-8Q^{2}}\in [1, 3]$, $r_{\min}$ is a simple root of $k^{2}+4P=0$, so the limit diverges. Therefore,
\begin{align}
\mathop{\mathrm{lim}}_{E \to E^{\pm}_{0}}\frac{r_{c}(E, k)-r_{\min}(k)}{E-E_{0}}=0,
\end{align}
which shows that the left and right derivatives of $r_{c}(E, k)$ at $E_{0}$ are both $0$; hence it is differentiable at that point, completing the proof of $C^{1}$ smoothness.

Using the fact that the critical curve satisfies $R=0$ and $\partial_{r}R=0$, we can directly obtain
\begin{align}
\pqty{\pdv{r_{c}}{E}}_{k}&=-\frac{2 \epsilon_{B}\sqrt{k^{2}+4P(r_{c})}}{r^{2}_{c}\partial^{2}_{r}R}, \label{pdvrcE}\\
\pqty{ \pdv{L_{c}}{E} }_{k} &=\frac{\pqty{E-\frac{k}{r_{c}}}r^{2}_{c}}{h(r_{c})L_{c}}, \label{pdvLcE}\\
\pqty{ \pdv{L_{c}}{k} }_{E} &=-\frac{r_{c}\pqty{E-\frac{k}{r_{c}}}}{h(r_{c})L_{c}}, \label{pdvLck}
\end{align}
where $\epsilon_{B}=+1$ for the $B_{+}$ branch and $\epsilon_{B}=-1$ for the $B_{-}$ branch; the critical point corresponds to a local minimum of $R(r)$, so $\partial^{2}_{r}R>0$; the future-directed condition of the physical momentum requires $E-k/r>0$, hence the signs of the above partial derivatives are evident.

We further introduce the critical impact parameter and its high-energy asymptotic value:
\begin{align}
b_{c}(E, k)& \coloneqq \frac{L_{c}(E, k)}{E}, \label{bcLbyE}\\
\mathop{\mathrm{lim}}_{E \to \infty}b_{c}(E,k)&= \mathop{\mathrm{lim}}_{E \to \infty}b_{c}(E,0)=\frac{r_{\text{ph}}}{\sqrt{h(r_{\text{ph}})}},  \label{limbcLbyE}
\end{align}
which means that ultrarelativistic particles behave like photons. Moreover, $b_{c}(E, k)$ is a bounded $C^{1}$ function for $E\in [1, +\infty)$.  For each fixed $E \geq 1$, the parametric expressions for $r_{c}(E, k)$ and $L_{c}(E, k)/E$ show that they are continuous in $k$ at $k=0$ (in the repulsive case, the $B_{\pm}$ branches reduce to the neutral one as $k\to 0$). Since $b_{c}(E, k)=L_{c}(E,k)/E$, it follows that
\begin{align}
\displaystyle\mathop{\mathrm{lim}}_{k \to 0}r_{c}(E, k) &=r_{c}(E, 0), & \displaystyle\mathop{\mathrm{lim}}_{k \to 0}b_{c}(E, k)&=b_{c}(E, 0). \label{limkrb}
\end{align}

\begin{figure*}[t]  
\begin{minipage}{0.48\textwidth}  
{ \centering \includegraphics[width=\linewidth]{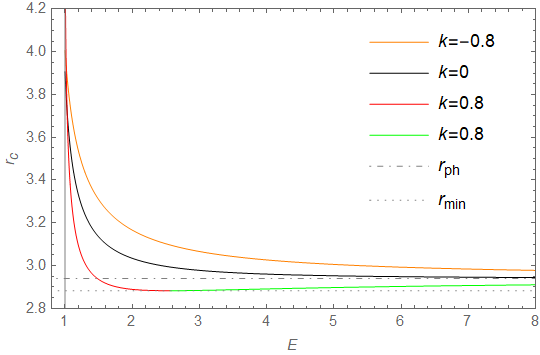}  }
\caption{Variation of the critical scattering point $r_{c}(E, k)$ with energy $E$ for different electromagnetic coupling parameters $k$ in RN spacetime. The plot parameter is $Q=0.3$.}
\label{fig:rcE}
\end{minipage}\hfill
\begin{minipage}{0.48\textwidth}
{ \centering  \includegraphics[width=\linewidth]{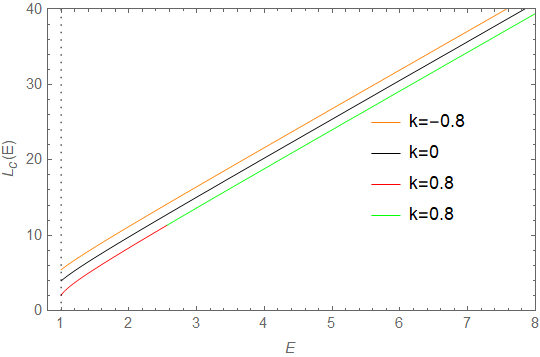} }
\caption{Variation of the critical angular momentum $L_{c}(E, k)$ with energy $E$ for different electromagnetic coupling parameters $k$ in RN spacetime. The plot parameter is $Q=0.3$.}
\label{fig:LcE}
\end{minipage}
\end{figure*}

Fig.~\ref{fig:rcE} shows the variation of $r_{c}(E, k)$ with energy $E$. For a fixed $k$, in the cases of electromagnetic attraction (orange line), charge neutrality (black line), and the low-energy branch of electromagnetic repulsion (red line), the position of the critical point decreases with increasing $E$ (all belonging to the $B_{+}$ branch); the behavior of the high-energy branch of electromagnetic repulsion (green line, belonging to the $B_{-}$ branch) is opposite to that of the low-energy branch. For different cases, the range of the critical point is the domain of definition of the respective $B_{\pm}$ branch.

The analysis of the tangency point between $E_{\text{eff}}$ and $U_{\text{eff}}$ indicates that for a given energy, the critical points for electromagnetic attraction and repulsion are located to the right and left of the maximum of the effective potential (i.e., the critical point for the charge-neutral case), respectively. This is reflected in the numerical curves, where the curves for $k=-0.8$ and $k=0.8$ lie above and below the $k=0$ curve, respectively.

Fig.~\ref{fig:LcE} is a numerical presentation of Eqs.~\eqref{pdvLcE}--\eqref{pdvLck}. For a fixed $k$, the critical angular momentum increases with $E$; for a fixed $E$, the required critical angular momentum gradually decreases as $k$ increases, so that the critical angular momenta for electromagnetic attraction and repulsion are, respectively, larger and smaller than that for the charge-neutral case.

At this point, we can analytically partition the absorption and scattering domains. At a point $x\in M$, we introduce a local angular parameter (non-conserved) $\sigma$ that relates the conserved axial angular momentum $L_{z}$ to the orbital angular momentum $L$:
\begin{align}
L_{z}=L\sin\theta_{x}\sin\sigma, \qquad \sigma\in [0,2\pi).
\end{align}
The physical momentum 1-form can then be uniquely expressed as
\begin{align}
p_{t}&=-E+\frac{k}{r_{x}}, & p_{r}&=\epsilon_{p} h(r_{x})^{-1}\sqrt{R(r_{x})}, & p_{\theta}&=L\cos\sigma, & p_{\varphi}&=L\sin\theta_{x}\sin\sigma, \label{p1-form}
\end{align}
where $\epsilon_{p}=-1$ for ingoing orbits and $\epsilon_{p}=+1$ for outgoing orbits. The coordinate parameters $\Bqty{E, L, \sigma}$ can be divided into the absorption domain $D_{\text{abs}}(x)$ and the scattering domain $D_{\text{scat}}(x)$:
\begin{align}
D_{\text{abs}}(x)&= [0,2\pi)_{\sigma}\times \Bqty{E \mid E \geq 1}\times D^{\text{abs}}_{L}(E); \label{Dabs}\\
D_{\text{scat}}(x)&= [0,2\pi)_{\sigma}\times \Bqty{E \mid r_{c}(E,k)\leq r_{x}}\times D^{\text{scat}}_{L}(E, r_{x}), \label{Dscat}
\end{align}
where, if $r_{c}(E, k)\leq r_{x}$ has no solution, it means that there is no scattering orbit passing through $r_{x}$. The physical momentum space at $x$ is partitioned into three mutually disjoint regions: the absorption region $\mathcal{A}_{x}$, the ingoing scattering region $\mathcal{S}^{\text{in}}_{x}$, and the outgoing scattering region $\mathcal{S}^{\text{out}}_{x}$, specifically
\begin{align}
\mathcal{A}_{x}& \equiv \Bqty{ p_{\mu} \mid D_{\text{abs}}(x), g^{t \mu}p_{\mu}>0,  p_{r}<0 },\\
\mathcal{S}^{\text{in}}_{x}& \equiv \Bqty{ p_{\mu} \mid D_{\text{scat}}(x), g^{t \mu}p_{\mu}>0, p_{r}<0 }, \label{Sinx}\\
\mathcal{S}^{\text{out}}_{x}& \equiv \Bqty{p_{\mu} \mid D_{\text{scat}}(x), g^{t \mu}p_{\mu}>0, p_{r}>0}. \label{Soutx}
\end{align}

\section{Review of the Vlasov theory foundations} \label{Sec3}

For a relativistic Vlasov gas whose constituent particles all have mass $m$, the distribution function $f$ is a non-negative function defined on the future mass shell (restoring the mass dimension)
\begin{align}
\Gamma^{+}_{m}\coloneqq \Bqty{ (x,p)\in T^{*}M \mid g^{-1}|_{x}(p, p) =-m^{2},\; g^{-1}(p, \cdot) \text{ is future-directed} },
\end{align}
and is denoted by $f(x^{\mu}, p_{\nu})$.

From the one-particle equation of motion Eq.~\eqref{equationmq} (or equivalently from Eq.~\eqref{KillingH}), the corresponding Vlasov equation can be derived:
\begin{align}
g^{\mu \nu}p_{\nu}\pdv{f}{x^{\mu}}-\frac{1}{2}p_{\alpha}p_{\beta}\pdv{g^{\alpha \beta}}{x^{\mu}}\pdv{f}{p_{\mu}}=q p_{\rho}F^{\rho}{}_{\sigma}\pdv{f}{p_{\sigma}}, \label{Vlasov-Maxwell}
\end{align}
a detailed derivation of which can be found in the literature~\cite{Acuna(2022),Liu(2026)}. On the other hand, by incorporating the symmetries induced by the spacetime symmetries (Killing vector fields) on the phase space, the distribution function can be further simplified to~\cite{Rioseco(2017a),Cieslik(2020)}
\begin{align}
f(x^{\mu}, p_{\nu})=f(E, L),
\end{align}
where $E$ and $L$ are the energy and orbital angular momentum of the particle, respectively.

Generally, on the spacetime $(M,g)$, one defines pointwise (at each $x\in M$) the particle current density $J_{\mu}$ and the stress-energy tensor $T_{\mu\nu}$, which are given by integrating the distribution function over the physical momentum space:
\begin{align}
J_{\mu}|_{x} &\coloneqq \int_{\Gamma^{+}_{m,x}} p_{\mu}f \dd{\mu_{x}}, \label{defJ} \\
 T_{\mu \nu}|_{x} &\coloneqq \int_{\Gamma^{+}_{m,x}} p_{\mu}p_{\nu}f \dd{\mu_{x}}, \label{defT}
\end{align}
where $\Gamma^{+}_{m, x}\subset T^{*}_{x}M$ is the restriction of the future mass shell to the fiber over $x$, and $\dd{\mu_{x}}$ is the invariant volume element on it, i.e., the induced volume element on $\Gamma^{+}_{m}$ from the natural volume element $\text{dvol}_{x}(p)=\sqrt{-\det(g^{\alpha \beta})}\dd^{4}p_{\mu}$ on $T^{*}_{x}M$. In the coordinates $\Bqty{\sigma, E, L}$,
\begin{align}
\dd{\mu_{x}}=\frac{1}{r^{2}}\frac{L}{\sqrt{R}}\dd{E}\dd{L}\dd{\sigma},
\end{align}
whose dimension is $[\dd{\mu_{x}}]=[m^{2}]$.

From the discussion in the previous section, $\Gamma^{+}_{m,x}=\mathcal{A}_{x} \sqcup \mathcal{S}^{\text{in}}_{x} \sqcup \mathcal{S}^{\text{out}}_{x}$, hence
\begin{align}
J^{\text{abs}}_{\mu}&=[m^{2}p_{\mu}]\int_{D_{\text{abs}}(x)} p^{\text{in}}_{\mu} \frac{f}{r^{2}}\frac{L}{\sqrt{R}}\dd{E}\dd{L}\dd{\sigma}, \label{Jmuabs}\\
J^{\text{scat}}_{\mu}&=[m^{2}p_{\mu}]\int_{D_{\text{scat}}(x)} (p^{\text{in}}_{\mu}+p^{\text{out}}_{\mu}) \frac{f}{r^{2}}\frac{L}{\sqrt{R}}\dd{E}\dd{L}\dd{\sigma}.\label{Jmuscat}
\end{align}
Here the dimensional factor $[m^{2}p_{\mu}]$ has been extracted, leaving dimensionless integrals. Since the boundaries of the parameter domain $\Bqty{E, L}$ are independent of $\sigma$, and the physical momentum 1-form (Eq.~\eqref{p1-form}) is used, the absorption and scattering parts of $J_{\theta}$ and $J_{\varphi}$ are both $0$, and $J^{\text{scat}}_{r}=0$. The integral expressions for $T_{\mu\nu}$ have a similar structure.

Using the Vlasov equation~\eqref{Vlasov-Maxwell} one can prove~\cite{Acuna(2022)} that for a charged Vlasov gas $(m, q)$, the pointwise-defined $J_{a}, T_{ab}$ satisfy the divergence relations
\begin{align}
\nabla^{a}J_{a}&=0, & \nabla^{a}T_{ab}&=-qJ_{c}\mathcal{F}^{c}{}_{b}, \label{graJT}
\end{align}
where $\mathcal{F}_{ab}$ denotes the total electromagnetic field, including the
self-field of the gas. After introducing the stress-energy tensor of the total electromagnetic field, $\mathcal{T}_{ab}=\frac{1}{4\pi}\pqty{\mathcal{F}_{ac}\mathcal{F}_{b}{}^{c}-\frac{1}{4}g_{ab}\mathcal{F}_{cd}\mathcal{F}^{cd} }$, the total stress-energy tensor $\mathscr{T}_{ab}=T_{ab}+\mathcal{T}_{ab}$ is divergence-free, i.e., $\nabla^{a}\mathscr{T}_{ab}=0$.

Since Eqs.~\eqref{defJ}--\eqref{defT} are all linear functionals of the distribution function, and the divergence relations~\eqref{graJT} are also linear, for a multicomponent gas $\Bqty{(m, q)_{\ell}, f_{\ell}}$ one can define
\begin{align}
J_{a}&\coloneqq \sum_{\ell}J^{(\ell)}_{a}, & \mathcal{J}_{a}&\coloneqq \sum_{\ell}q_{\ell}J^{(\ell)}_{a}, & T_{ab}&\coloneqq \sum_{\ell}T^{(\ell)}_{ab},\label{totalJJT}
\end{align}
which represent the total particle current density, the total charge current density, and the total stress-energy tensor of the multicomponent gas, respectively, and satisfy
\begin{align}
\nabla^{a}J_{a}&=0, & \nabla^{a}\mathcal{J}_{a}&=0, \label{graJ} \\
 \nabla^{a}T_{ab}&=-\mathcal{J}_{c}\mathcal{F}^{c}{}_{b}, & \nabla^{a}\mathcal{F}_{ab}&=-4\pi \mathcal{J}_{b}. \label{graT}
\end{align}

For a plasma Vlasov gas with weak electric coupling ($k\to 0$), it follows from the  relations~\eqref{limkrb} that, under the pure background field assumption, the collective behavior of the multicomponent plasma reduces to that of the neutral gas~\cite{Cieslik(2020)}. In such a case, the specific study of a two-component plasma $(m, \pm q)$ can be found in the literature~\cite{Liu(2026)}.

\section{Physical analysis of the plasma} \label{Sec4}

\subsection{Distribution function of the particle source}

Consider a spherically symmetric particle source at infinity that fuels the black hole accretion~\cite{Rioseco(2017a)}. After fixing the particle source distribution function $f_{\infty}$, the distribution function can be extended to finite positions via the Vlasov equation~\eqref{Vlasov-Maxwell}. In the following discussion, we focus only on the case where the distribution is energy-dependent, i.e., $f_{\infty}=f_{\infty}(E)$.

The simplest model is a monoenergetic distribution, assuming that any component $\ell$ of the gas satisfies
\begin{align}
f_{\ell \infty}\equiv \beta_{\ell} \delta(E-\epsilon_{\ell}),
\end{align}
where $\beta_{\ell}$ is a normalization coefficient. The distribution describes a continuous emission of particles with a fixed energy $\epsilon_{\ell}$ from infinity or provides an approximate description of an energy spectrum concentrated near $\epsilon_{\ell}$, i.e., $E\in [\epsilon_{\ell}, \epsilon_{\ell}+\Delta \epsilon_{\ell}],\; \Delta \epsilon_{\ell}/\epsilon_{\ell}\ll 1$.

If the particle source at infinity is a stationary isothermal perfect gas in local thermal equilibrium, it is assumed that any component $\ell$ obeys the J\"uttner distribution:
\begin{align}
f_{\ell \infty}  \equiv \alpha_{\ell} e^{- \frac{E}{k_{B}T_{\ell}}}=\alpha_{\ell}e^{-z_{\ell} E}, \label{funJ}
\end{align}
where $\alpha_{\ell}$ is a normalization coefficient, $k_{B}$ and $T_{\ell}$ are the Boltzmann constant and the temperature parameter, respectively. The second expression has been made dimensionless; denoting the particle mass of component $\ell$ by $m_{\ell}$, $E \leftarrow E/m_{\ell}$, and $z_{\ell}\equiv m_{\ell}/(k_{B}T_{\ell})$.

In this paper, we only consider a two-component plasma system $\Bqty{(m_{1}, q_{1}), (m_{2}, q_{2}); q_{1}q_{2}<0}$. Let the particle number density (defined in Eq.~\eqref{defn}) of each component at infinity be $n_{\ell \infty}$, and the total density be $n_{\infty}$, then the plasma provides the boundary conditions
\begin{align}
n_{\infty}&=n_{1\infty}+n_{2\infty}, & 0&=q_{1}n_{1\infty}+q_{2}n_{2\infty}.
\end{align}

We define the asymptotic weight of component $\ell$ as
\begin{align}
\lambda_{\ell} \coloneqq \frac{n_{\ell \infty}}{n_{\infty}},
\end{align}
which can be equivalently expressed through the charge ratio or the ratio of the electromagnetic coupling parameters, i.e.,
\begin{align}
\lambda_{1}&\coloneqq \frac{n_{1 \infty}}{n_{1\infty}+n_{2\infty}}=\frac{q_{2}}{q_{2}-q_{1}}=\frac{m_{2}k_{2}}{m_{2}k_{2}-m_{1}k_{1}}, \label{deflambda}
\end{align}
and $\lambda_{2}$ is obtained by swapping the subscripts $1 \leftrightarrow 2$; clearly $\lambda_{1}+\lambda_{2}=1$.

\subsection{Particle current density vector and particle number accretion rate}

First, we give the nonzero components of the particle current density vector (after integrating over $\sigma$ and $L$):
\begin{align}
 J^{\text{abs}}_{t}&=- \frac{2\pi m^{3}}{h(r)}\int^{\infty}_{1}f_{\infty}(E)(E-k/r)\pqty{  R^{1/2}_{0}-R^{1/2}_{c} }\dd{E},\\
 J^{\text{scat}}_{t}&=-  \frac{4\pi m^{3}}{h(r)} \int^{\infty}_{1}\chi_{E,r} f_{\infty}(E) (E-k/r)R^{1/2}_{c}\dd{E},\\
 J^{\text{abs}}_{r}&=- \frac{ \pi m^{3} }{r^{2} h(r)}\int^{\infty}_{1} f_{\infty}(E)  L^{2}_{c}(E, k) \dd{E},
\end{align}
where $R_{0}\equiv R|_{L=0}$, $R_{c}\equiv R|_{L=L_c(E, k)}$. $\chi_{E, r}$ is the energy characteristic function of the scattering part at position $r$: $\chi_{E, r}=1$ if $r_{c}(E, k) \leq r$, and $0$ otherwise. Note that $r_{\text{mb}}(k)$ is a finite value, so as $r\to \infty$, $\chi_{E, r}=1$. The asymptotic expansions of the above integrals are
\begin{align}
J^{(\ell)\text{abs}}_{t}& \sim -\frac{\pi m^{3}}{r^{2}} \int^{\infty}_{1}f_{\infty}(E)	E \frac{L^{2}_{c}(E, k)}{\sqrt{E^{2}-1}} \dd{E},\\
J^{(\ell)\text{scat}}_{t}& \sim -4\pi m^{3} \int^{\infty}_{1}f_{\infty}(E)  \bqty{E\sqrt{E^{2}-1}+\frac{(2E^{2}-1)(E-k)}{r\sqrt{E^{2}-1}}} \dd{E}, \label{Jtscatinf}\\
J^{(\ell)\text{abs}}_{r}& \sim - \frac{\pi m^{3} }{r^{2}}\int^{\infty}_{1}f_{\infty}(E)  L^{2}_{c}(E, k) \dd{E}.\label{Jrabsinf}
\end{align}

Define the particle number density of component $\ell$ as
\begin{align}
n_{\ell} \coloneqq \sqrt{-g(J^{(\ell)},J^{(\ell)})},  \label{defn}
\end{align}
the particle number density at any position can be obtained using the integral formulas; clearly $n_{\ell}=n_{\ell}(r)$. At infinity,
\begin{align}
 n_{\ell \infty}\equiv n_{\ell}(r)|_{r \to \infty}=-J^{\text{scat}}_{t}|_{r \to \infty},
\end{align}
which yields, for the monoenergetic and J\"uttner distributions, respectively,
\begin{align}
n_{\ell\infty}&=4\pi m^{3} \beta_{\ell} \epsilon_{\ell} \sqrt{\epsilon^{2}_{\ell}-1}, & \beta_{\ell} m^{3}&=\frac{n_{\ell\infty}}{4\pi \epsilon_{\ell}\sqrt{\epsilon^{2}_{\ell}-1}},\\
n_{\ell\infty}&= 4\pi m^{3} \alpha_{\ell}  \frac{ K_{2}(z_{\ell})}{z_{\ell}}, & 
\alpha_{\ell}m^{3} &= \frac{n_{\ell \infty} z_{\ell}}{4\pi K_{2}(z_{\ell})},
\end{align}
where $K_{2}(z)$ is the modified Bessel function of the second kind.

For the total particle current density vector $J=J^{(1)}+J^{(2)}$ of the plasma, the corresponding particle number density $n$ satisfies
\begin{align}
n^{2}=(n_{1}+n_{2})^{2}+2n_{1}n_{2}\cdot \pqty{\cosh\Delta \psi-1},\label{nplasma}
\end{align}
where $\Delta \psi=\psi_{1}-\psi_{2}$, and $\psi_{\ell}$ is the rapidity of the particle current density vector of component $\ell$ with respect to a (fictitious) neutral-gas reference frame, defined by
\begin{align}
\cosh \pqty{\Delta \psi}  \equiv -g\pqty{\frac{J^{(1)}}{n_{1}}, \frac{J^{(2)}}{n_{2}} },  \label{crossn}
\end{align}
a detailed derivation of which is given in Appendix~\ref{AppA}, where a numerical study of $\frac{(\Delta \psi)^{2}}{2}$ at the horizon is also presented. Table~\ref{tab:data} lists the maximum values of $\eval{\frac{(\Delta \psi)^{2}}{2}}_{\max}$ at the horizon for typical electromagnetic coupling parameters.

\begin{table}[htbp]
\begin{ruledtabular}
\begin{tabular}{c | c c  c | c c c}
& $Q=0.33$ & $Q=0.66$ & $Q=0.99$ & $Q=0.33$ & $Q=0.66$ & $Q=0.99$  \\
\hline
	$k=\pm 0.33$ & $0.67\%$ & $0.90\%$ & $3.38\%$ & $0.18\%$ & $0.23\%$ & $0.85\%$  \\
	$k=\pm 0.66$ & $2.96\%$ & $4.04\%$ & $18.50\%$ & $0.72\%$ & $0.94\%$ & $3.61\%$  \\
	$k=\pm 0.99$ & $7.96\%$ & $11.29\%$ & $100.00\%$ & $1.68\%$ & $2.21\%$ & $9.05\%$  \\
\end{tabular}
\end{ruledtabular}
\caption{The first row gives the black hole charge (specific charge) $Q$, and the first column lists the electromagnetic coupling parameters $k_{1}, k_{2}$ of the plasma. The table entries are the values of $\eval{\frac{(\Delta \psi)^{2}}{2}}_{\max}$ at the horizon for different electromagnetic coupling cases. The left block corresponds to a cold plasma with $\lg(\epsilon-1)\sim -2$ (or $\lg(z)\sim 2$); the right block corresponds to a plasma at finite temperature with $\lg(\epsilon-1)\sim 0$ (or $\lg(z)\sim 0$).}
\label{tab:data}
\end{table}

In the subsequent study we do not consider cold plasmas in the limit of extreme electromagnetic coupling. From the typical values in Table~\ref{tab:data} it can be seen that at the horizon $\frac{(\Delta \psi)^{2}}{2}$ can always be treated as a small quantity. The electromagnetic coupling parameter $k$ measures the deviation of the charged gas from the neutral gas, and the closer the gas is to the black hole, the more pronounced the electromagnetic effect is, so this deviation is maximal near the horizon. Hence $\frac{(\Delta \psi)^{2}}{2}$ is universally small outside the horizon as well. Using this property, Eq.~\eqref{nplasma} can be rewritten as
\begin{align}
n & \simeq n_{1}+n_{2}+\frac{n_{1}n_{2}}{n_{1}+n_{2}}\cdot \frac{(\Delta \psi)^{2}}{2}, \label{equ:n}\\
\delta_{n}&\equiv \frac{n-(n_{1}+n_{2})}{n_{1}+n_{2}}\simeq \frac{n_{1}n_{2}}{(n_{1}+n_{2})^{2}}\frac{(\Delta \psi)^{2}}{2} \leq \frac{(\Delta \psi)^{2}}{8}.
\end{align}

We call $\frac{n_{\ell}}{n_{\ell\infty}}$ the compression factor of the particle number density of component $\ell$, which describes the change of the gas density from infinity to a finite radius. Taking the leading order of Eq.~\eqref{equ:n}, the compression factor of the plasma is related to those of the components by
\begin{align}
\frac{n}{n_{\infty}} &\simeq  \lambda_{1}\frac{n_{1}}{n_{1\infty}}+\lambda_{2}\frac{n_{2}}{n_{2\infty}}. \label{nlambdann}
\end{align}
That is, the plasma compression factor is, to high accuracy, the weighted average of the compression factors of the components, with the asymptotic weights $\lambda_{\ell}$; the relative error is controlled by $\mathscr{O}\bigl((\Delta \psi)^{2}\bigr)$.

For a cold plasma, the corresponding analytic result can be obtained directly with the help of Eq.~\eqref{Laplace}.

Consider the divergence of the particle current density vector:
\begin{align}
\nabla_{a}J^{a}=\pdv{J^{t}}{t}+\frac{1}{r^{2}}\pdv{r^{2}J^{r}}{r}+\frac{1}{\sin\theta}\pdv{\sin\theta J^{\theta}}{\theta}+\pdv{J^{\varphi}}{\varphi}\equiv 0,
\end{align}
so the particle current density of each component is a conserved current. The particle number accretion rate of the black hole can be defined with the help of a two-dimensional topological sphere on an arbitrary spacelike hypersurface. Choosing a two-sphere $S_{r}$ on a hypersurface of constant $t$, we have
\begin{align}
\frac{\dot{\mathcal{N}}^{(\ell)}}{n_{\ell\infty}} &\coloneqq -\frac{M^{2}}{n_{\ell\infty}}\int_{S_{r}} J^{(\ell)r}r^{2}\sin\theta \dd{\theta}\dd{\varphi}=-\frac{4\pi M^{2}}{n_{\ell\infty}}\pqty{r^{2}J^{(\ell)r}}\notag\\
&= \frac{ 4\pi^{2} m^{3}M^{2} }{n_{\ell\infty}}\int^{\infty}_{1} f_{\infty}(E) E^{2} b^{2}_{c}(E, k) \dd{E}\notag\\
&=\begin{cases}
\pi M^{2} \frac{\epsilon_{\ell}}{\sqrt{\epsilon^{2}_{\ell}-1}} b^{2}_{c}(\epsilon_{\ell}, k_{\ell}), & f_{\infty}\propto \delta(E-\epsilon_{\ell}),\\
\pi  M^{2} \frac{ z_{\ell}}{ K_{2}(z_{\ell})}\int^{\infty}_{1}e^{-z_{\ell} E} E^{2}  b^{2}_{c}(E, k_{\ell}) \dd{E}, & f_{\infty}\propto e^{-z_{\ell}E},
\end{cases}\label{defdotN}
\end{align}
where the negative sign ensures that $\dot{\mathcal{N}}^{(\ell)}$ is the particle number flowing into the black hole; $M^{2}$ gives the dimension of the integration volume element on $S_{r}$, and $b_{c}(E, k)\equiv \frac{L_{c}(E,k)}{E}$ is the critical impact parameter defined earlier (see Eq.~\eqref{bcLbyE}). 

The analytical expression for the particle number accretion rate $\frac{\dot{\mathcal{N}}^{(\ell)}}{n_{\ell\infty}}$ shows that, for a given distribution function, a single-component gas satisfies
\begin{align}
\pdv{k}\pqty{ \frac{\dot{\mathcal{N}}^{(\ell)}}{n_{\ell\infty}} }=\pi M^{2} \int^{\infty}_{1}\bqty{\frac{4\pi m^{3}}{n_{\ell\infty}}f_{\infty}(E)}\partial_{k}L^{2}_{c}(E, k) \dd{E} <0, \label{particalkN}
\end{align}
where the last step uses $\partial_{k}L_{c}(E, k)<0$ (see Eq.~\eqref{pdvLck}).

Expressing the total particle current density $J=J^{(1)}+J^{(2)}$ in terms of the asymptotic weights,
\begin{align}
 \frac{J}{n_{\infty}}&=\lambda_{1}\frac{J^{(1)}}{n_{1\infty}}+\lambda_{2} \frac{J^{(2)}}{n_{2\infty}}, \label{Jlambda}
\end{align}
one immediately finds that the normalized particle accretion rate of the plasma satisfies the strict weighted-average relation
\begin{align}
\frac{\dot{\mathcal{N}}}{n_{\infty}}&=\lambda_{1}\frac{\dot{\mathcal{N}}^{(1)}}{n_{1\infty}}+\lambda_{2}\frac{\dot{\mathcal{N}}^{(2)}}{n_{2\infty}}. \label{linefunN}
\end{align}

The charge accretion rate $\dot{\mathcal{Q}}$ is defined as
\begin{align}
\dot{\mathcal{Q}}&\coloneqq q_{1}\dot{\mathcal{N}}^{(1)}+q_{2}\dot{\mathcal{N}}^{(2)}=q_{1}n_{1\infty}\pqty{\frac{\dot{\mathcal{N}}^{(1)}}{n_{1\infty}} - \frac{\dot{\mathcal{N}}^{(2)}}{n_{2\infty}} }.\label{defdotQ}
\end{align}
From Eq.~\eqref{particalkN}, for a two-component plasma, when the distribution functions of the two species are identical (i.e., $\epsilon_1=\epsilon_2$ or $z_1=z_2$), accretion inevitably drives the RN black hole toward a Schwarzschild black hole (the black hole tends to become charge-neutral). When the distribution functions differ, the evolution can only be determined through specific parameters $\epsilon_{\ell}$ or $z_{\ell}$ (by numerical integration).

After an appropriate normalization, the total charge accretion rate $\dot{\mathcal{Q}}\equiv \sum_{\ell} q_{\ell} \dot{\mathcal{N}}^{(\ell)}$ and the charge accretion rates of the individual components $\dot{\mathcal{Q}}^{(\ell)}\equiv q_{\ell} \dot{\mathcal{N}}^{(\ell)}$ also form a weighted-average relation,
\begin{align}
\frac{\dot{\mathcal{Q}}}{\bar{q}n_{\infty}}&=\lambda_{1}\frac{\dot{\mathcal{Q}}^{(1)}}{q_{1}n_{1 \infty}}+\lambda_{2}\frac{\dot{\mathcal{Q}}^{(2)}}{q_{2}n_{2 \infty}},\label{linefunQ}\\
\bar{q} & \equiv  \frac{\dot{\mathcal{N}}^{(1)}}{\dot{\mathcal{N}}}q_{1}+\frac{\dot{\mathcal{N}}^{(2)}}{\dot{\mathcal{N}}}q_{2}, \label{defbarq}
\end{align}
where $\bar{q}$ has the physical meaning of the average charge of all accreted particles.

The particle number accretion rate, Eq.~\eqref{defdotN}, depends only on the boundary parameters and the critical impact parameter $b_{c}(E, k)$ (or equivalently $L_{c}(E, k)$). For the J\"uttner distribution, considering the extreme gas limits, i.e., the high-temperature limit $z \to 0^{+}$ and the low-temperature limit $z \to +\infty$, we have
\begin{align}
\frac{z}{K_{2}(z)}\int^{\infty}_{1}e^{-z E}E^{2} b^{2}_{c}(E, k)\dd{E} 
\simeq \begin{cases}
b^{2}_{c}(\infty , 0), & z \to 0^{+},\\
\sqrt{\frac{2z}{\pi}}b^{2}_{c}(1, k), & z\to +\infty,
\end{cases}
\end{align}
which shows that the extreme gas cases under the J\"uttner distribution correspond to monoenergetic gases with $\epsilon \to +\infty$ and $\epsilon \to 1^{+}$, respectively. The following asymptotic formulas were used in the calculation:
\begin{align}
 K_{n}(z) |_{z \to 0} & \simeq \frac{\Gamma(n)}{2} \pqty{ \frac{2}{z} }^{n},\\
K_{n}(z) |_{z \to \infty} & \simeq \sqrt{\frac{\pi}{2 z}}e^{-z}\pqty{ 1+ \frac{4n^{2}-1}{8z}+o(z^{-1}) }. \label{Bessel}\\
	 \int^{\infty}_{1}e^{-z x}x^{\gamma}\mathcal{B}(x) \dd{x} &\simeq \begin{cases}
\frac{\Gamma(\gamma+1)}{z^{\gamma+1}}\mathcal{B}(\infty), & z \to 0^{+} , \gamma >-1,\\
\frac{e^{-z}}{z}\mathcal{B}(1), & z \to +\infty,
\end{cases} \label{Laplace} 
\end{align}
where the Laplace-type integral in Eq.~\eqref{Laplace} requires that $\mathcal{B}(x)$ is smooth at $x=1$ and has a limit as $x\to +\infty$. Our $b^{2}_{c}(E, k)$ satisfies this property with respect to $E$, and in the high-temperature limit we used Eq.~\eqref{limbcLbyE}, i.e., $b_{c}(\infty, k)=b_{c}(\infty,0)$. 

The analysis of the extreme gas cases, especially the high-temperature limit $z\to 0^{+}$, shows that in the background electromagnetic field of the RN spacetime, the behavior of high-energy particles tends to that of photons; however, at finite temperatures, $b_{c}(E, k)$ deviates from $b_{c}(E, 0)$, so the electromagnetic coupling parameter $k$ mainly affects accretion at finite temperatures.

\subsection{Stress-energy tensor of the matter field}

To study key information such as the energy density and principal pressures of the gas, it is convenient to decompose the stress-energy tensor according to its eigenvectors. The total stress-energy tensor of the mixed gas is the sum of the contributions of the components, i.e., $T_{ab}=\sum_{\ell}T^{(\ell)}_{ab}$. Below we list the nonzero components of $T^{(\ell)}_{ab}$ for each component in the coordinate basis; for brevity the component index $\ell$ is omitted:
\begin{align}
	T^{(\text{abs})}_{tt}&=\frac{2\pi m^{4}}{h(r)}\int^{\infty}_{1}f_{\infty}(E)(E-k/r)^{2} \pqty{R^{1/2}_{0}-R^{1/2}_{c}}\dd{E},\\
	T^{(\text{scat})}_{tt}&=\frac{4\pi m^{4}}{h(r)}\int^{\infty}_{1}\chi_{E, r} f_{\infty}(E)(E-k/r)^{2} R^{1/2}_{c}\dd{E},\\
	T^{(\text{abs})}_{rr}&=\frac{2\pi m^{4}}{3h(r)^{3}}\int^{\infty}_{1}f_{\infty}(E)\pqty{R^{3/2}_{0}-R^{3/2}_{c}}\dd{E},\\
	T^{(\text{scat})}_{rr}&=\frac{4\pi m^{4}}{3h(r)^{3}}\int^{\infty}_{1}\chi_{E, r}f_{\infty}(E)R^{3/2}_{c}\dd{E},
\end{align}
\begin{align}
	T^{(\text{abs})}_{\theta \theta}&=\frac{\pi m^{4}M^{2}}{r^{2}}\int^{\infty}_{1}f_{\infty}(E)\bqty{\mathcal{I}_{3}(0)-\mathcal{I}_{3}(L_{c})}\dd{E},\\
	T^{(\text{scat})}_{\theta \theta}&=\frac{2\pi m^{4}M^{2}}{r^{2}}\int^{\infty}_{1} \chi_{E, r} f_{\infty}(E)\mathcal{I}_{3}(L_{c})\dd{E},\\
	T_{\varphi \varphi }&=\sin^{2}\theta T_{\theta \theta},\\
	T^{(\text{abs})}_{rt}&=\frac{\pi m^{4}}{r^{2}h(r)}\int^{\infty}_{1}f_{\infty}(E)(E-k/r)L^{2}_{c}(E, k)\dd{E}, \label{Tgrt}
\end{align}
where the expression for $\mathcal{I}_{3}(L)$ is
\begin{align}
\mathcal{I}_{3}(L)&=\frac{\sqrt{R(L)}}{3C^{4}}(3A^{2}-R(L)), & A^{2} &\equiv (E-k/r)^{2}-h(r), & C^{2}&\equiv h(r)/r^{2}.
\end{align}

The nonzero components of $T_{ab}$ in the static observer's frame are
\begin{align}
T_{00}&=T_{tt}/h, &T_{11}&=hT_{rr}, &T_{22}&=T_{33}=T_{\theta\theta}/r^{2},&  T_{01}&=T_{rt}. \label{defTij}
\end{align}

Let $\Bqty{w_{\alpha}, \alpha=0,1,2,3}$ be the eigenvectors of $T^{a}{}_{b}$, i.e., $T^{a}{}_{b}(w_{\alpha})^{b} \propto (w_{\alpha})^{a}$. In the dual frame $\Bqty{w^{\alpha}}$ the stress-energy tensor can be written as
\begin{align}
T&=\varepsilon w^{0}\otimes w^{0}+p_{\text{rad}}w^{1} \otimes w^{1}+p_{\text{tan}}\pqty{ w^{2} \otimes w^{2}+w^{3}\otimes w^{3}},
\end{align}
where $\varepsilon>0,\; p_{\text{rad}},\; p_{\text{tan}}$ are the proper energy density, proper radial pressure, and proper tangential pressure in the eigenframe, respectively. Their explicit expressions are
\begin{align}
\varepsilon &=\frac{D+(T_{00}-T_{11}) }{2}, & p_{\text{rad}}&=\frac{D-(T_{00}-T_{11}) }{2}, & p_{\text{tan}} &=T_{22},
\end{align}
with $D \equiv \sqrt{(T_{00}+T_{11})^{2}-4T^{2}_{01}}$.

Asymptotic analysis of $T_{\hat{\alpha}\hat{\beta}}$ shows that as $r\to\infty$ the only nonzero components are
\begin{align}
T_{00}&=4\pi m^{4}\int^{\infty}_{1}f_{\infty}(E) E^{2}\sqrt{E^{2}-1}\dd{E},\\
T_{11}&=T_{22}=T_{33}=\frac{4\pi m^{4}}{3}\int^{\infty}_{1}f_{\infty}(E)\pqty{E^{2}-1}^{3/2}\dd{E},
\end{align}
where the factor $M^{2}$ in the expression for $T_{\theta \theta}$ has been canceled by $r^{2}$ in Eq.~\eqref{defTij}.

At spatial infinity the energy density is $\varepsilon_{\ell \infty}=T_{00}$. For the monoenergetic and J\"uttner distributions we obtain, respectively,
\begin{align}
\varepsilon_{\ell\infty}&=4\pi m^{4}\beta_{\ell} \epsilon^{2}_{\ell}\sqrt{\epsilon^{2}_{\ell}-1}, & \beta_{\ell}m^{4}&=\frac{ \varepsilon_{\ell \infty}}{4\pi \epsilon^{2}_{\ell}\sqrt{\epsilon^{2}_{\ell}-1} },\\
\varepsilon_{\ell \infty}&=4\pi m^{4}\alpha_{\ell}\bqty{\frac{K_{1}(z_{\ell})}{z_{\ell}}+3\frac{K_{2}(z_{\ell})}{z^{2}_{\ell}}}, & \alpha_{\ell}m^{4}&=\frac{ \varepsilon_{\ell\infty}\, z^{2}_{\ell}}{4\pi \bqty{ z_{\ell}K_{1}(z_{\ell})+3K_{2}(z_{\ell})}}.
\end{align}
The average single-particle energy of each component is
\begin{align}
\hat{ \varepsilon}_{\ell \infty} \coloneqq \frac{ \varepsilon_{\ell \infty}}{m_{\ell} n_{\ell \infty}}=
\begin{cases}
\epsilon_{\ell}, & f_{\infty}\propto \delta(E-\epsilon_{\ell}),\\
 \frac{z_{\ell}K_{1}(z_\ell)+3K_{2}(z_{\ell})}{z_{\ell}K_{2}(z_{\ell})}, & f_{\infty}\propto e^{-z_{\ell}E}.
\end{cases}
\end{align}
For the extreme gas cases under the J\"uttner distribution: as $z_{\ell} \to +\infty$, $\hat{ \varepsilon}_{\ell \infty}=m_{\ell}\pqty{1+\frac{3}{2z}}=m_{\ell}+\frac{3}{2}k_{B}T_{\ell}$; as $z_{\ell} \to 0^{+}$, $\hat{ \varepsilon}_{\ell \infty}=m_{\ell}\frac{3}{z}=3k_{B}T_{\ell}$. The low-temperature limit corresponds to a monatomic perfect gas ($T_{\ell}=0$ corresponds to dust), and the high-temperature limit corresponds to a radiation gas (photon gas).

The principal pressure is $p=p_{\text{rad}}=p_{\text{tan}}=T_{11}$, and
\begin{align}
\frac{p_{\ell \infty}}{ \varepsilon_{\ell\infty}}&=\begin{cases}
\frac{1}{3}\pqty{ 1- \frac{1}{ \epsilon^{2}_{\ell}}}\in \pqty{0, \frac{1}{3}}, & f_{\infty} \propto \delta(E- \epsilon_{\ell}),\\
\pqty{3+\frac{z_{\ell} K_{1}(z_{\ell})}{K_{2}(z_{\ell})}}^{-1}\in \pqty{0, \frac{1}{3}}, & f_{\infty}\propto e^{-z_{\ell}E}.
\end{cases}
\end{align}
When $\epsilon_{\ell}\to 1^{+}$ or $z_{\ell} \to +\infty$, $p_{\ell\infty}/\varepsilon_{\ell \infty}=0$, corresponding to a dust equation of state; when $\epsilon_{\ell} \to +\infty$ or $z_{\ell} \to 0^{+}$, $p_{\ell\infty}/\varepsilon_{\ell \infty}=1/3$, corresponding to a radiation gas equation of state.

At infinity both the energy density and the principal pressure are linear functions of the stress-energy tensor. Hence for the plasma $T=T^{(1)}+T^{(2)}$ we have
\begin{align}
\varepsilon_{\infty} &=\varepsilon_{1\infty}+\varepsilon_{2\infty}, & p_{\infty}=p_{1\infty}+p_{2\infty}.
\end{align}
Introducing the asymptotic weight of the energy density,
\begin{align}
\Lambda_{\ell} \coloneqq \frac{ \varepsilon_{\ell\infty}}{ \varepsilon_{\infty}}, \label{defLambda}
\end{align}
the total stress-energy tensor can be expressed as
\begin{align}
\frac{T}{ \varepsilon_{\infty}}&=\Lambda_{1} \frac{T^{(1)}}{ \varepsilon_{1\infty}}+\Lambda_{2} \frac{T^{(2)}}{ \varepsilon_{2\infty}}. \label{linefunT}
\end{align}

Since $p_{\text{tan}}$ is a linear function of the stress-energy tensor, the total tangential pressure satisfies
\begin{align}
p_{\text{tan}}&=p_{1\text{tan}}+p_{2\text{tan}},
\end{align}
and the weighted-average relation
\begin{align}
\frac{p_{\text{tan}}}{ \varepsilon_{\infty}}&=\Lambda_{1}\frac{p_{1\text{tan}}}{ \varepsilon_{1\infty}}+\Lambda_{2} \frac{p_{2\text{tan}}}{ \varepsilon_{2\infty}}. \label{linefunptan}
\end{align}

For the energy density and radial pressure, the standard analysis in Appendix~\ref{AppB} gives (see Eqs.~\eqref{Appepsilon}--\eqref{Appprad})
\begin{align}
\varepsilon  &\simeq \pqty{\varepsilon_{1}+\varepsilon_{2}}+\frac{(\varepsilon_{1}+p_{1\text{rad}})(\varepsilon_{2}+p_{2\text{rad}})}{ \varepsilon_{1}+p_{1\text{rad}}+\varepsilon_{2}+p_{2\text{rad}}}\cdot (\Delta \vartheta)^{2}, \\
	 p_{\text{rad}}&\simeq \pqty{p_{1\text{rad}}+p_{2\text{rad}}}+\frac{(\varepsilon_{1}+p_{1\text{rad}})(\varepsilon_{2}+p_{2\text{rad}})}{ \varepsilon_{1}+p_{1\text{rad}}+\varepsilon_{2}+p_{2\text{rad}}}\cdot (\Delta \vartheta)^{2},
\end{align}
where $\Delta \vartheta=\vartheta_{1}-\vartheta_{2}$, and $\vartheta_{\ell}$ is the rapidity of the timelike eigenvector of component $\ell$ with respect to a (fictitious) neutral-gas reference frame. In analogy with the smallness of $\frac{(\Delta \psi)^{2}}{2}$, as long as we are not dealing with a cold plasma in the limit of extreme electromagnetic coupling, $(\Delta \vartheta)^{2}$ can be regarded as a small quantity. Consequently, in close analogy with the compression factor for the particle number density, we obtain at finite radii
\begin{align}
\frac{ \varepsilon}{ \varepsilon_{\infty}}&\simeq \Lambda_{1}\frac{ \varepsilon_{1}}{ \varepsilon_{1\infty}}+\Lambda_{2}\frac{ \varepsilon_{2}}{ \varepsilon_{2\infty}},\label{varepsilonsimeq}\\
\frac{ p_{\text{rad}}}{ \varepsilon_{\infty}}&\simeq \Lambda_{1}\frac{ p_{1\text{rad}}}{ \varepsilon_{1\infty}}+\Lambda_{2}\frac{ p_{2\text{rad}} }{ \varepsilon_{2\infty}}.\label{pradsimeq}
\end{align}

Having discussed the particle current density vector and the stress-energy tensor of the Vlasov gas in detail, we note that the Vlasov gas becomes a perfect fluid if $J^{a}$ is an eigenvector of $T^{a}{}_{b}$ and the pressure is isotropic, i.e., if
\begin{align}
\frac{T_{tt}+T^{rr}-2T^{r}{}_{t}}{T_{tt}+T^{rr}+2T^{r}{}_{t}}&=\pqty{\frac{J_{t}-J^{r}}{J_{t}+J^{r}}}^{2}, & p_{\text{rad}}&=p_{\text{tan}},
\end{align}
where the first condition follows from Eqs.~\eqref{PsilogJ} and \eqref{varthetalogT} in the appendices. In general, a Vlasov gas does not satisfy these two conditions, and therefore it is not a perfect fluid.

\subsection{Energy accretion rate}

For the divergence-free total stress-energy tensor $\nabla^{a}\mathscr{T}_{ab}=0$, analogous to the particle number accretion rate~\eqref{defdotN}, one can define the energy accretion rate
\begin{align}
\dot{\mathcal{E}}&=4\pi r^{2}\mathscr{T}^{r}{}_{t}=4\pi r^{2} \pqty{ T^{r}{}_{t}+\mathcal{T}^{r}{}_{t}}\notag\\
&=\bqty{ 4\pi^{2}m^{4}M^{2}\int^{\infty}_{1}f_{\infty}(E)E L^{2}_{c}(E, k)\dd{E} }^{(\text{M})} \notag\\
& \qquad + \bqty{ 4\pi r^{2}\mathcal{T}^{r}{}_{t} -\frac{4\pi^{2}m^{4}M^{2} k}{r}\int^{\infty}_{1}f_{\infty}(E)L^{2}_{c}(E, k)\dd{E} }^{(\text{EM})}\notag\\
&\equiv \dot{\mathcal{E}}^{(\text{M})}+\dot{\mathcal{E}}^{(\text{EM})}. 
\end{align}
Since the stress-energy tensor $T_{ab}$ constructed from the physical momentum $p_{a}=\pi_{a}-qA_{a}$ satisfies (Eq.~\eqref{graJT}) $\nabla^{a}T_{ab}=-qJ_{c}\mathcal{F}^{c}{}_{b}$, the quantity $4\pi r^{2}T^{r}{}_{t}$ alone cannot serve as a well-defined energy accretion rate. However, through the decomposition of the total divergence-free stress-energy tensor $\mathscr{T}_{ab}=T_{ab}+\mathcal{T}_{ab}$, we obtain an $r$-independent piece $\dot{\mathcal{E}}^{(\text{M})}$ and an $r$-independent piece $ \dot{\mathcal{E}}^{(\text{EM})}$, with
\begin{align}
\dot{\mathcal{E}}^{(\text{M})}=\displaystyle\mathop{\mathrm{lim}}_{r \to \infty}4\pi r^{2}T^{r}{}_{t}.
\end{align}
In the following we focus on $\dot{\mathcal{E}}^{(\text{M})}$ and interpret it as the asymptotic value of the matter-field energy accretion rate. The electromagnetic interaction contribution $ \dot{\mathcal{E}}^{(\text{EM})}$ should be studied within a fully self-consistent Vlasov--Maxwell system, rather than in the present test-particle simplification, which suffers from electromagnetic divergences analogous to the neglect of self-gravity in a
neutral test-particle model.

Define the energy accretion rate $\dot{\mathcal{E}}^{(\ell)}$ in the test field $F_{\text{RN}}$:
\begin{align}
\frac{\dot{\mathcal{E}}^{(\ell)}}{ \varepsilon_{\ell\infty}}& \coloneqq \frac{4\pi^{2}m^{4}M^{2}}{\varepsilon_{\ell\infty}}\int^{\infty}_{1}f_{\infty}(E)E L^{2}_{c}(E, k_{\ell})\dd{E}\notag\\
&=\frac{4\pi^{2}m^{4}M^{2}}{\varepsilon_{\ell\infty}}\int^{\infty}_{1}f_{\infty}(E)E^{3} b^{2}_{c}(E, k_{\ell})\dd{E}\notag\\
&=\begin{cases}
\pi M^{2}\frac{\epsilon_{\ell}}{\sqrt{\epsilon^{2}_{\ell}-1}}b^{2}_{c}(\epsilon_{\ell},k_{\ell}), & f_{\infty} \propto \delta(E-\epsilon_{\ell}),\\
\pi M^{2} \frac{z^{2}_{\ell}}{z_{\ell}K_{1}(z_{\ell})+3K_{2}(z_{\ell})} \int^{\infty}_{1}e^{-z_{\ell} E} E^{3} b^{2}_{c}(E, k_{\ell})\dd{E}, & f_{\infty} \propto e^{-z_{\ell} E}.
\end{cases}
\end{align}
Similar to the particle number accretion rate, the energy accretion rate depends only on the boundary conditions, where the dimension $M^{2}$ comes from the area element $4\pi r^{2}$. Comparing with the particle number accretion rate, for a monoenergetic distribution or an extreme J\"uttner distribution ($z \to 0^{+}$ or $z \to +\infty$), we have $\dot{\mathcal{E}}^{(\ell)}/\varepsilon_{\ell\infty}=\dot{\mathcal{N}}^{(\ell)}/n_{\ell\infty}$.

The total energy accretion rate is
\begin{align}
\frac{\dot{\mathcal{E}}}{ \varepsilon_{\infty}}=\Lambda_{1} \frac{\dot{\mathcal{E}}^{(1)}}{ \varepsilon_{1\infty}}+\Lambda_{2} \frac{\dot{\mathcal{E}}^{(2)}}{ \varepsilon_{2\infty}}. \label{linefunE}
\end{align}

\section{Weighted-average relations and their classification}\label{Sec5}

In the preceding sections we have examined in detail the various physical quantities of the accreting plasma Vlasov gas. All of them are built from the particle current density vector $J=J^{(1)}+J^{(2)}$ and the stress-energy tensor $T=T^{(1)}+T^{(2)}$. After a suitable normalization, the total quantities exhibit an underlying structure of a weighted average of the contributions from the individual components, namely
\begin{align}
 \frac{J}{n_{\infty}}&=\lambda_{1}\frac{J^{(1)}}{n_{1\infty}}+\lambda_{2} \frac{J^{(2)}}{n_{2\infty}}, \tag{\ref{Jlambda}}\\
 \frac{T}{ \varepsilon_{\infty}}&=\Lambda_{1} \frac{T^{(1)}}{ \varepsilon_{1\infty}}+\Lambda_{2} \frac{T^{(2)}}{ \varepsilon_{2\infty}},\tag{\ref{linefunT}}
\end{align}
where the contribution of each component $\ell$ is a linear functional of the distribution function $f_{\infty}$. The weights $\lambda_{\ell}$ and $\Lambda_{\ell}$ are, respectively, the asymptotic weight of the particle number density (Eq.~\eqref{deflambda}) and that of the energy density (Eq.~\eqref{defLambda}).

The macroscopic quantities of the gas and the accretion rates of the black hole can be divided into two classes according to whether or not they are linear functionals of $f_{\infty}$.

The class of linear functionals includes the tangential pressure $p_{\text{tan}}$, the particle number accretion rate $\dot{\mathcal{N}}$, the charge accretion rate $\dot{\mathcal{Q}}$, and the energy accretion rate $\dot{\mathcal{E}}$. For these quantities the relation between the total value and the component values is a strict weighted average:
\begin{align}
\frac{p_{\text{tan}}}{ \varepsilon_{\infty}}&=\Lambda_{1}\frac{p_{1\text{tan}}}{ \varepsilon_{1\infty}}+\Lambda_{2}\frac{p_{2\text{tan}}}{ \varepsilon_{2\infty}}, \tag{\ref{linefunptan}}\\
\frac{\dot{\mathcal{N}}}{n_{\infty}}&=\lambda_{1}\frac{\dot{\mathcal{N}}^{(1)}}{n_{1\infty}}+\lambda_{2}\frac{\dot{\mathcal{N}}^{(2)}}{n_{2\infty}}, \tag{\ref{linefunN}}\\
\frac{\dot{\mathcal{Q}}}{\bar{q}n_{\infty}}&=\lambda_{1}\frac{\dot{\mathcal{Q}}^{(1)}}{q_{1}n_{1 \infty}}+\lambda_{2}\frac{\dot{\mathcal{Q}}^{(2)}}{q_{2}n_{2 \infty}},\tag{\ref{linefunQ}}\\
\frac{\dot{\mathcal{E}}}{ \varepsilon_{\infty}}&=\Lambda_{1} \frac{\dot{\mathcal{E}}^{(1)}}{ \varepsilon_{1\infty}}+\Lambda_{2} \frac{\dot{\mathcal{E}}^{(2)}}{ \varepsilon_{2\infty}}. \tag{\ref{linefunE}}
\end{align}

The class of nonlinear functionals comprises the particle number density $n$, the energy density $\varepsilon$, and the radial pressure $p_{\text{rad}}$. For these quantities the relation between the whole and the parts is more complicated. However, provided one does not consider a cold plasma in the limit of extreme electromagnetic coupling, the total quantities still satisfy the weighted-average relations to high accuracy:
\begin{align}
\frac{n}{n_{\infty}} &\simeq  \lambda_{1}\frac{n_{1}}{n_{1\infty}}+\lambda_{2}\frac{n_{2}}{n_{2\infty}}, \tag{\ref{nlambdann}}\\
\frac{ \varepsilon}{ \varepsilon_{\infty}}&\simeq \Lambda_{1}\frac{ \varepsilon_{1}}{ \varepsilon_{1\infty}}+\Lambda_{2}\frac{ \varepsilon_{2}}{ \varepsilon_{2\infty}},\tag{\ref{varepsilonsimeq}}\\
\frac{ p_{\text{rad}}}{ \varepsilon_{\infty}}&\simeq \Lambda_{1}\frac{ p_{1\text{rad}}}{ \varepsilon_{1\infty}}+\Lambda_{2}\frac{ p_{2\text{rad}} }{ \varepsilon_{2\infty}}.\tag{\ref{pradsimeq}}
\end{align}

In summary, after normalization the macroscopic physical quantities of a plasma Vlasov gas uniformly exhibit a structure of a weighted average of the contributions of the individual components. The only distinction is that for the linear functionals the relations are exact equalities, whereas for the nonlinear functionals they are approximations whose accuracy is controlled by $\frac{(\Delta\psi)^2}{2}$ or $(\Delta\vartheta)^2$. This classification provides a clear theoretical framework for understanding the collective behavior of a two-component collisionless plasma.

\section{Numerical Results}\label{Sec6}

We now present numerical illustrations of the analytical results obtained in Secs.~\ref{Sec4} and~\ref{Sec5} and test the weighted-average structure. When a physical quantity $A(u)$ is plotted as a function of a parameter $u$, the horizontal axis is taken as $x=\log u$ (or $x=\lg u$) and the vertical axis is labeled by $A$; the numerical curve $A$--$x$ should then be understood as $A(e^{x})$ (or $A(10^{x})$). In the following three subsections we investigate successively the particle number density $n$; the energy density $\varepsilon$ and the principal pressures $p_{\text{rad}}$, $p_{\text{tan}}$; and the particle number accretion rate $\dot{\mathcal{N}}$ and the energy accretion rate $\dot{\mathcal{E}}$.

The corresponding charge and mass accretion rates follow directly from $\dot{\mathcal{N}}$ and the particle charges (or masses).

\subsection{Particle number density}

We first examine the radial profile of $J_{t}=J^{(\text{abs})}_{t}+J^{(\text{scat})}_{t}$, the only nontrivial component of the particle current density, since $J_{\theta}=J_{\varphi}=0$ and $J^{r}\propto 1/r^{2}$.

\begin{figure*}[htbp]
\begin{minipage}{0.48\textwidth}  
{ \centering \includegraphics[width=\linewidth]{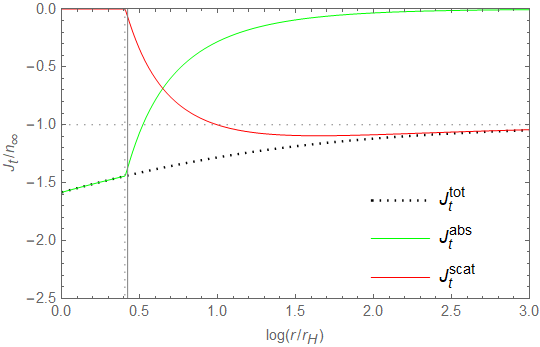} }
\caption{Particle current density component $J_{t}$ as a function of $x\equiv\log(r/r_{H})$ for a single charged gas with electromagnetic repulsion. 
Parameters: $Q=0.66$, $k=0.66$, $z=1$.}
\label{fig:Jtkp}
\end{minipage}\hfill
\begin{minipage}{0.48\textwidth}
{ \centering  \includegraphics[width=\linewidth]{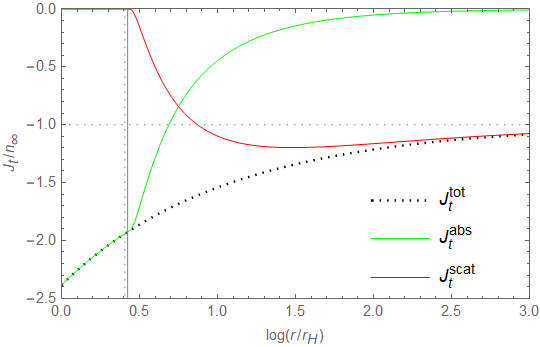} }
\caption{Particle current density component $J_{t}$ as a function of $x\equiv\log(r/r_{H})$ for a single charged gas with electromagnetic attraction. Parameters: $Q=0.66$, $k=-0.66$, $z=1$.}
\label{fig:Jtkm}
\end{minipage}
\end{figure*}

Figs.~\ref{fig:Jtkp} and~\ref{fig:Jtkm} show $J_{t}$ for a single-component gas with electromagnetic repulsion ($k=0.66$) and attraction ($k=-0.66$). For the total current, $|J^{\text{tot}}_{t}|$ is larger in the attractive case. The scattering part $J^{\text{scat}}_{t}$ vanishes below $r_{\min}(k)$ for repulsion (dotted gray line) and below $r_{\text{ph}}$ for attraction (solid gray line); the small deviation of the numerical vanishing points from these theoretical values arises from the high-energy tail of the distribution $f_{\infty}\propto e^{-zE}$ together with the critical curve $r_{c}(E)$ (see Fig.~\ref{fig:rcE}), which is not fully resolved in the numerical integration.

\begin{figure*}[htbp]
\begin{minipage}{0.48\textwidth}  
{ \centering \includegraphics[width=\linewidth]{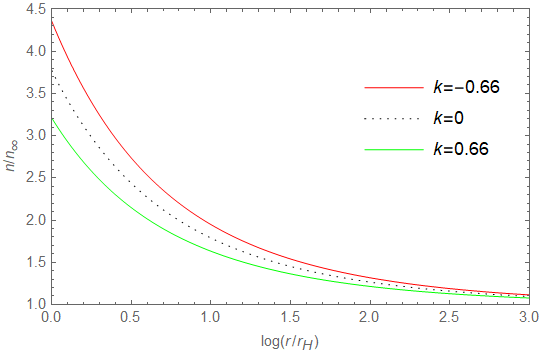} }
\caption{Particle number density $n$ as a function of $x\equiv\log(r/r_{H})$ for a single-component gas. 
Parameters: $Q=0.66$, $z=1$.}
\label{fig:singnr}
\end{minipage}
\hfill
\begin{minipage}{0.48\textwidth}
{ \centering  \includegraphics[width=\linewidth]{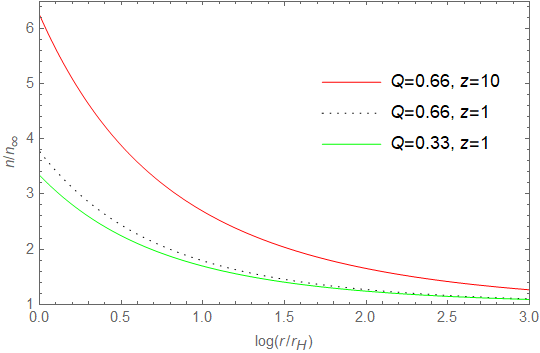} }
\caption{Particle number density $n$ as a function of $x$ for a neutral gas ($k=0$), showing the dependence on black hole charge $Q$ and the parameter $z$.}
\label{fig:singnr2}
\end{minipage}
\end{figure*}

Figs.~\ref{fig:singnr} and~\ref{fig:singnr2} show the particle number density $n$ as a function of $x$ for a single-component gas. In all cases $n$ decreases outward. Fig.~\ref{fig:singnr} shows that, at fixed $x$, electromagnetic attraction (repulsion) enhances (suppresses) $n$ relative to the neutral case, with the deviation growing near the horizon. Fig.~\ref{fig:singnr2} shows that increasing either the black hole charge $Q$ or the parameter $z$ increases the compression factor $n/n_{\infty}$. Note that when comparing different $Q$, the same $x$ corresponds to different physical radii ($r=r_H e^x$), while varying $z$ also changes $n_\infty$ (for a fixed normalization factor $\alpha$).

\begin{figure*}[htbp]
\begin{minipage}{0.48\textwidth}  
{ \centering \includegraphics[width=\linewidth]{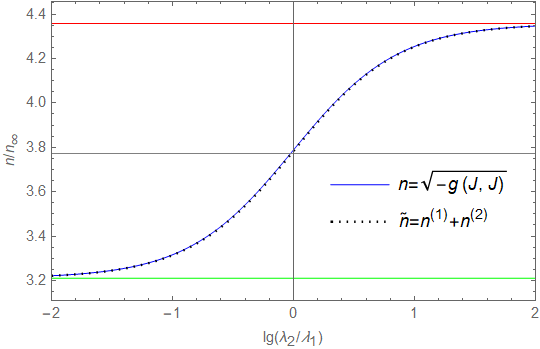} }
\caption{Proper particle number density of the plasma at the horizon as a function of the asymptotic weight ratio $\lambda_2/\lambda_1$. 
Parameters: $Q=0.66$, $k=\pm 0.66$, $z=1$.}
\label{fig:totnr}
\end{minipage}
\hfill
\begin{minipage}{0.48\textwidth}
{ \centering  \includegraphics[width=\linewidth]{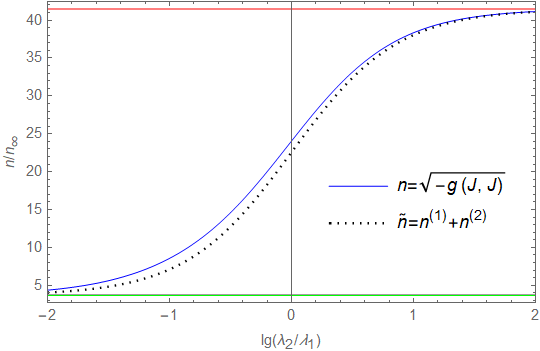} }
\caption{Proper particle number density of the plasma at the horizon as a function of the asymptotic weight ratio $\lambda_2/\lambda_1$ for an extremely cold plasma. Parameters: $Q=0.99$, $k=\pm 0.99$, $z=30$. The curves illustrate the breakdown of the weighted-average approximation for certain ratios.}
\label{fig:coldplasma}
\end{minipage}
\end{figure*}

Figs.~\ref{fig:totnr} and~\ref{fig:coldplasma} test the weighted-average relation \eqref{nlambdann} for the particle number density at the horizon. We compare the exact plasma density $n$ from Eq.~\eqref{nplasma} with the algebraic sum $\tilde{n}=n_{1}+n_{2}$, normalized as
\begin{align}
\frac{\tilde{n}}{n_{\infty}}=\lambda_{1}\frac{n_{1}}{n_{1\infty}}+\lambda_{2}\frac{n_{2}}{n_{2\infty}}.
\end{align}
For symmetric couplings ($k_{1}+k_{2}=0$), the neutrality condition $\lambda_{1}q_{1}+\lambda_{2}q_{2}=0$ implies $\lambda_{1}m_{1}=\lambda_{2}m_{2}$, fixing the relation between the horizontal-axis ratio and the mass ratio.

In Fig.~\ref{fig:totnr} the curves of $\frac{n}{n_{\infty}}$ and $\frac{\tilde{n}}{n_{\infty}}$ almost coincide, confirming Eq.~\eqref{nlambdann}. In Fig.~\ref{fig:coldplasma}, however, they deviate significantly for certain $\lambda_{2}/\lambda_{1}$, indicating failure of the weighted-average approximation in a cold plasma with extreme electromagnetic coupling. For the parameters $Q=0.99$, $k=\pm0.99$, $z=30$, one has $\Delta\psi=1.2625$. Defining $a_{1}\equiv \frac{n_{1}}{n_{1\infty}}\simeq 3.68$, $a_{2}\equiv \frac{n_{2}}{n_{2\infty}}\simeq 41.49$, and $v\equiv\lambda_{2}/\lambda_{1}$, the absolute error follows from Eq.~\eqref{nplasma} as
\begin{align}
\Delta(u)&=\frac{n}{n_{\infty}}-\frac{\tilde{n}}{n_{\infty}}\notag\\
&=(1+u)^{-1}\Bigl[\sqrt{(a_{1}+a_{2}v)^{2}+2a_{1}a_{2}(\cosh\Delta\psi-1)v}-(a_{1}+a_{2}v)\Bigr],
\end{align}
which is well approximated by $\Delta(v) \sim 1.85\,\mathrm{sech}^{2}[1.06(\lg v+0.483)]$.

In this subsection we have used a controlled-variable approach to study the influence of various initial conditions on the particle number density $n$. For instance, when only $k$ differs, at the same position $x$ the particle number density for electromagnetic attraction (repulsion) is larger (smaller) than that of the neutral gas; such a qualitative conclusion often holds also for $\varepsilon$, $p_{\text{rad}}$, and $p_{\text{tan}}$. Therefore, in the next subsection we will not repeat the same exhaustive controlled-variable tests.

\subsection{Radial profiles of energy density and principal pressures}

In this subsection we numerically display the radial profiles of the compression factors of the energy density and principal pressures, $\frac{\varepsilon}{\varepsilon_{\infty}}, \frac{p_{\text{rad}}}{\varepsilon_{\infty}}, \frac{p_{\text{tan}}}{\varepsilon_{\infty}}$, as well as the corresponding per-particle averages $\frac{\varepsilon}{mn},\frac{p_{\text{rad}}}{mn}, \frac{p_{\text{tan}}}{mn}$, as functions of the position $x \equiv \log(r/r_{H})$, where $m$ and $n$ are the particle mass and particle number density, respectively.

\begin{figure*}[htbp]
\begin{minipage}{0.48\textwidth}  
{ \centering \includegraphics[width=\linewidth]{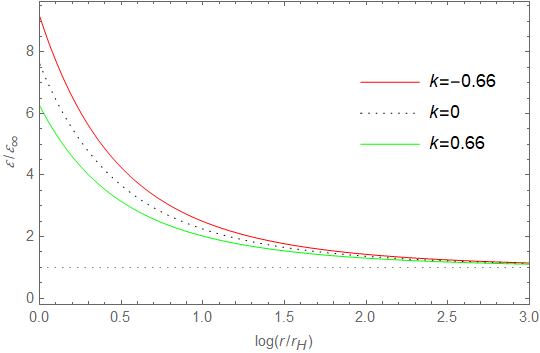} }
\caption{Proper energy density $\varepsilon$ as a function of $x\equiv\log(r/r_{H})$ for different electromagnetic couplings. 
Parameters: $Q=0.66$, $z=1$.}
\label{fig:varepsilonr}
\end{minipage}
\hfill
\begin{minipage}{0.48\textwidth}
{ \centering  \includegraphics[width=\linewidth]{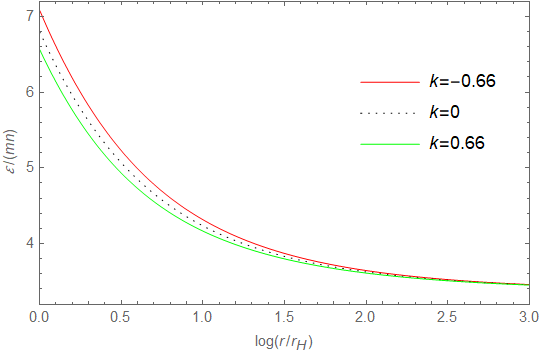} }
\caption{Average energy per particle $\varepsilon/(m n)$ as a function of $x$, for the same parameter set.}
\label{fig:varepsilonr2}
\end{minipage}
\end{figure*}

\begin{figure*}[htbp]
\begin{minipage}{0.48\textwidth}  
{ \centering \includegraphics[width=\linewidth]{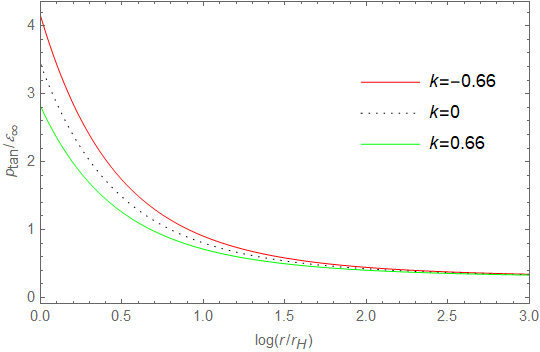} }
\caption{Tangential pressure $p_{\mathrm{tan}}$ as a function of $x\equiv\log(r/r_{H})$. 
Parameters: $Q=0.66$, $z=1$.}
\label{fig:ptanr}
\end{minipage}
\hfill
\begin{minipage}{0.48\textwidth}
{ \centering  \includegraphics[width=\linewidth]{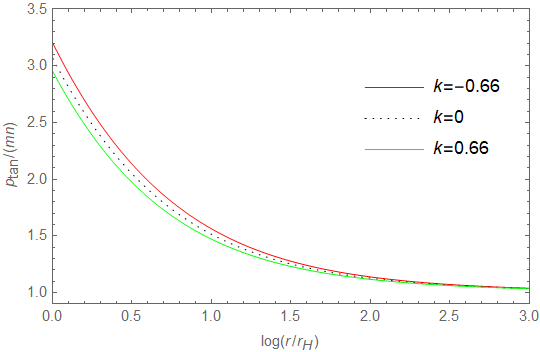} }
\caption{Average tangential pressure per particle $p_{\mathrm{tan}}/(m n)$ as a function of $x$, for the same parameter set.}
\label{fig:ptanr2}
\end{minipage}
\end{figure*}

Figs.~\ref{fig:varepsilonr} and~\ref{fig:varepsilonr2} show, respectively, the energy density and the average energy per particle as functions of the position $x$. The numerical results indicate that both quantities decrease with increasing $x$, and that the values for electromagnetic attraction (repulsion) are larger (smaller) than the corresponding values for the neutral gas; the deviation from the neutral gas becomes more pronounced as one approaches the horizon. Figs.~\ref{fig:ptanr} and~\ref{fig:ptanr2} present the numerical curves for the tangential pressure, $\frac{p_{\text{tan}}}{\varepsilon_{\infty}}$ and $\frac{p_{\text{tan}}}{mn}$; their overall behavior is similar to that of the energy density.

\begin{figure*}[htbp]
\begin{minipage}{0.48\textwidth}  
{ \centering \includegraphics[width=\linewidth]{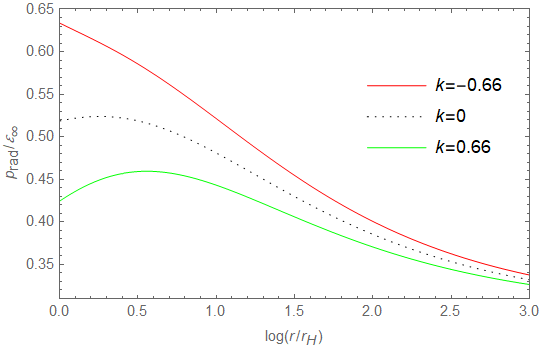} }
\caption{Radial pressure $p_{\mathrm{rad}}$ as a function of $x\equiv\log(r/r_{H})$. 
Parameters: $Q=0.66$, $z=1$.}
\label{fig:pradr}
\end{minipage}
\hfill
\begin{minipage}{0.48\textwidth}
{ \centering  \includegraphics[width=\linewidth]{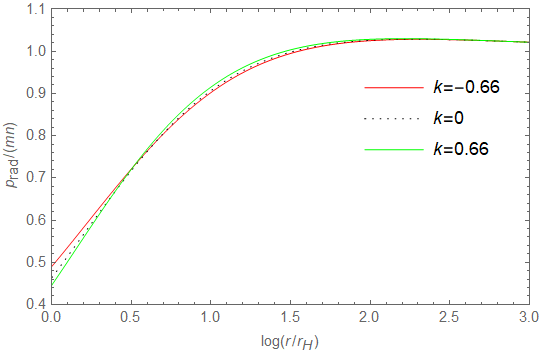} }
\caption{Average radial pressure per particle $p_{\mathrm{rad}}/(m n)$ as a function of $x$, for the same parameter set.}
\label{fig:pradr2}
\end{minipage}
\end{figure*}

The behavior of the radial pressure $\frac{p_{\text{rad}}}{\varepsilon_{\infty}}$ shown in Fig.~\ref{fig:pradr} is somewhat more intricate, yet it still obeys the basic property that the values for electromagnetic attraction (repulsion) exceed (fall below) those for the neutral gas. For the per-particle average radial pressure, Fig.~\ref{fig:pradr2}, the curves for different cases almost overlap at moderate radii and become smaller closer to the black hole; they separate again near the horizon and satisfy the same property: attraction $>$ neutral $>$ repulsion.

Comparing the radial and tangential pressures, Fig.~\ref{fig:pradr} versus Fig.~\ref{fig:ptanr} (or Fig.~\ref{fig:pradr2} versus Fig.~\ref{fig:ptanr2}), one sees that near the horizon the radial pressure is significantly smaller than the tangential pressure. This is one manifestation of how a Vlasov gas differs from a perfect fluid. Ref.~\cite{Rioseco(2017a)} explicitly computed the radial and tangential pressures at the horizon in the low-temperature limit ($z \to +\infty$) and pointed out that the Vlasov gas accretion model provides a partial explanation for why the actual accretion rate is much lower than the Bondi accretion rate.

\subsection{Accretion Rates}

We now study the dependence of the normalized accretion rates $\dot{\mathcal{N}}/n_{\infty}$ and $\dot{\mathcal{E}}/\varepsilon_{\infty}$ on the black hole charge $Q$ (all charges are dimensionless, cf. Sec.~\ref{Sec2}). Since the plasma accretion rates are exact weighted averages of the component rates (Sec.~\ref{Sec5}), it suffices to consider a single component with a fixed particle charge $q$; varying $Q$ then scans the coupling $k=qQ$.

\begin{figure*}[htbp]
\begin{minipage}{0.48\textwidth}
{ \centering \includegraphics[width=\linewidth]{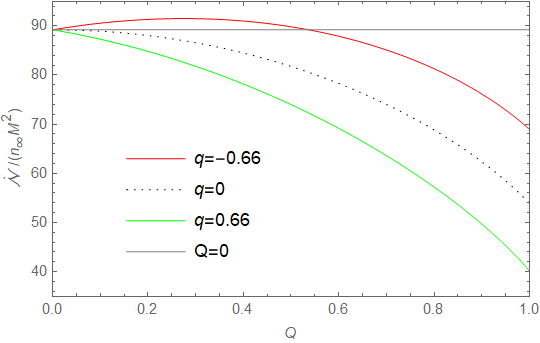} }
\caption{Particle number accretion rate $\dot{\mathcal{N}}/(n_{\infty}M^{2})$ as a function of $Q$ for $z=1$.}
\label{fig:dotN}
\end{minipage}
\hfill
\begin{minipage}{0.48\textwidth}
{ \centering  \includegraphics[width=\linewidth]{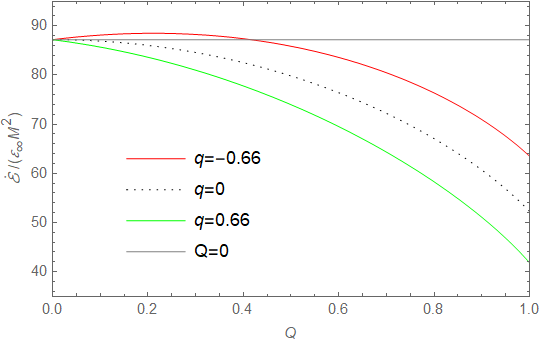} }
\caption{Energy accretion rate $\dot{\mathcal{E}}/(\varepsilon_{\infty}M^{2})$ as a function of $Q$ for $z=1$.}
\label{fig:dotE}
\end{minipage}
\end{figure*}

Fig.~\ref{fig:dotN} shows the particle number accretion rate for a fixed gas (fixed $q$) as a function of the black hole charge $Q$. It can be seen that, for a given $Q$, the accretion rate for electromagnetic attraction ($k<0$) is larger than that for the neutral case ($k=0$), whereas the rate for electromagnetic repulsion ($k>0$) is smaller. For the neutral and electromagnetically repulsive cases, $\dot{\mathcal{N}}/n_{\infty}$ decreases monotonically with increasing $Q$; for the electromagnetically attractive case, $\dot{\mathcal{N}}/n_{\infty}$ first increases monotonically and then decreases monotonically as $Q$ grows. In particular, at $Q=0$ the curve for the neutral case has a horizontal tangent, while in the presence of electromagnetic interaction the tangent is no longer horizontal. Theoretically, we have
\begin{align}
\eval{\dv{(\dot{\mathcal{N}}/n_{\infty})}{Q }}_{Q=0}  \propto \int^{\infty}_{1}f_{\infty}(E)L_{c}(E, 0)\bqty{ \eval{\pdv{L_{c}(E,k)}{Q}}_{Q=0}+ \eval{\pdv{L_{c}(E, k)}{k}}_{k=0}\cdot q }\dd{E},
\end{align}
where $\propto$ denotes equality up to a positive proportionality factor. When evaluating the partial derivatives $\partial_{Q}$ and $\partial_{k}$, $k$ is treated as an implicit function of $Q$. For the neutral gas the tangent at $Q=0$ is horizontal, so $\eval{\partial_{Q}L_{c}(E,k)}_{Q=0}=0$; moreover, Eq.~\eqref{pdvLck} gives $\eval{\partial_{k} L_{c}(E, k)}_{k=0}<0$, which explains the behavior of the tangents of the accretion-rate curves at $Q=0$ in the different cases.

The numerical curve for the energy accretion rate $\dot{\mathcal{E}}/\varepsilon_{\infty}$ (Fig.~\ref{fig:dotE}) exhibits almost exactly the same behavior as the particle number accretion rate, the only difference originating from the fact that the ratio of the normalization factors $\varepsilon_{\infty}/n_{\infty}$ is generally not equal to~$1$.

\begin{figure*}[htbp]
\begin{minipage}{0.48\textwidth}
{ \centering \includegraphics[width=\linewidth]{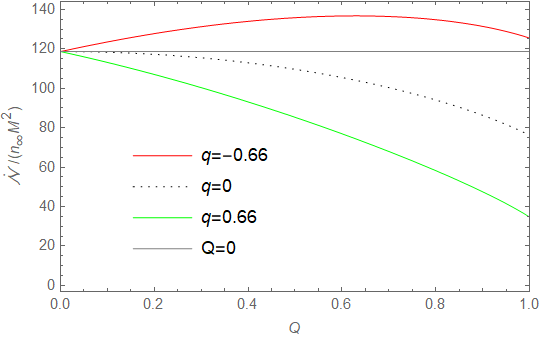} }
\caption{Particle number accretion rate $\dot{\mathcal{N}}/(n_{\infty}M^{2})$ as a function of $Q$ for $z=5$.}
\label{fig:dotN2}
\end{minipage}
\hfill
\begin{minipage}{0.48\textwidth}
{ \centering  \includegraphics[width=\linewidth]{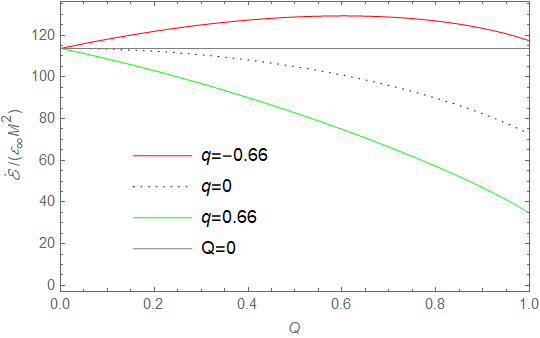} }
\caption{Energy accretion rate $\dot{\mathcal{E}}/(\varepsilon_{\infty}M^{2})$ as a function of $Q$ for $z=5$.}
\label{fig:dotE2}
\end{minipage}
\end{figure*}

Figs.~\ref{fig:dotN2}--\ref{fig:dotE2} present the accretion-rate curves for a relatively low temperature ($z=5$). A comparison with the $z=1$ case shows that the monotonicity properties of the individual curves are similar, but the actual numerical values differ considerably. In particular, the accretion rate of the electromagnetically attractive component is rather sensitive to the parameter $z$ (when $z \to +\infty, \sqrt{z}$ diverges).

These numerical results are consistent with the charge evolution discussed in Sec.~\ref{Sec4}: identical distributions drive the black hole toward neutrality, whereas differing distributions require specific parameter choices to determine the sign of the charge accretion.

\section{Conclusion}\label{Sec7}

We have constructed a complete analytical framework for the steady-state, spherically symmetric accretion of a two-component plasma Vlasov gas in the Reissner--Nordstr\"om spacetime.

We obtained analytical (in parametric form) $C^1$ expressions for the critical angular momentum, critical radius, and critical impact parameter as functions of the energy $E$, and showed that in the limit $k \to 0$ they reduce to their neutral-gas counterparts. This provides a rigorous basis for partitioning the physical phase space into absorption and scattering domains and gives, for the first time, a complete analytical characterization of the accretion and scattering behavior of charged particles in a background electromagnetic field.

For both monoenergetic and J\"uttner distributions, we further computed explicit integral expressions for the particle number density, energy density, and anisotropic pressures. Electromagnetic attraction ($k<0$) enhances these macroscopic quantities and the associated accretion rates, whereas electromagnetic repulsion ($k>0$) suppresses them, clearly illustrating the regulatory role of electromagnetic interactions in the accretion process.

For a two-component plasma, we identified a weighted-average structure: after normalization, the tangential pressure, particle number accretion rate, and energy accretion rate are exactly equal to the weighted averages of the corresponding contributions from the individual components. Except for cold plasmas with extreme electromagnetic coupling, the compression factors of the particle number density, energy density, and radial pressure also satisfy the same weighted-average relations to high accuracy. The weighting factors $\lambda_\ell$ and $\Lambda_\ell$ are determined solely by the particle source at infinity.

Numerical results confirm the above weighted-average relations and show that near the horizon the radial pressure is much smaller than the tangential pressure, exhibiting strong anisotropy that is markedly different from a perfect fluid. For a single-component gas, the accretion rate as a function of the black hole charge $Q$ decreases monotonically in the electromagnetically repulsive and charge-neutral cases, while in the electromagnetically attractive case it generally first increases and then decreases. An analysis of the charge accretion rate shows that when the distribution functions of the two components are identical, the Reissner--Nordstr\"om black hole naturally evolves toward the Schwarzschild limit; when they differ, the direction of charge evolution is determined by the specific parameters through numerical integration.

\begin{acknowledgments}

This work is supported by the National Natural Science Foundation of China (NSFC) under Grant nos.~12235019,~12275106.

\end{acknowledgments}

\appendix

\section{Rapidity of the charged gas}\label{AppA}

For the convenience of the discussions in this appendix and in Appendix~\ref{AppB}, we list here the tetrad of the static observer in RN spacetime, $\{e_{\alpha}\}$, and its dual $\{e^{\alpha}\}$:
\begin{align}
e_{0}&=h^{-1/2}\partial_{t}, & e_{1}&=h^{1/2}\partial_{r}, & e_{2}&=r^{-1}\partial_{\theta}, & e_{3}&=(r\sin\theta)^{-1}\partial_{\varphi},\\
e^{0}&=h^{1/2}\dd{t}, & e^{1}&=h^{-1/2}\dd{r}, & e^{2}&=r\dd{\theta}, &e^{3}&=r\sin\theta \dd{\varphi}.
\end{align}

Let the future-directed unit timelike vector $\hat{J}(k_{\ell})\equiv J(k_{\ell})/n_{\ell}$ have the rapidity $\Psi(k)$ relative to the static observer $e_{0}$:
\begin{align}
\hat{J}(k) &=  \cosh\Psi(k) e_{0}+  \sinh\Psi(k) e_{1}.
\end{align}
Obviously $\hat{J}(0)$ corresponds to the comoving observer of the neutral gas. We can therefore express $\hat{J}(k)$ as
\begin{align}
\hat{J}(k) &=\cosh\psi(k) U+  \sinh\psi(k) V,\\
U &\equiv \cosh\Psi(0) e_{0}+  \sinh\Psi(0) e_{1} =\hat{J}(0),\\
V &\equiv  \sinh\Psi(0) e_{0}+ \cosh\Psi(0) e_{1},
\end{align}
where $\psi(k)$ is the relative rapidity of $\hat{J}(k)$ with respect to the neutral-gas reference frame, defined by
\begin{align}
\psi(k) \coloneqq \Psi(k)-\Psi(0).
\end{align}
A direct calculation yields the result required in Eq.~\eqref{crossn}:
\begin{align}
-g\pqty{ \hat{J}(k_{1}), \hat{J}(k_{2}) }=\cosh\pqty{\psi(k_{1})-\psi(k_{2})}.
\end{align}

From the expression for $\hat{J}(k)$ we can solve for the rapidity $\Psi(k)$:
\begin{align}
\tanh\Psi(k)=\frac{g\pqty{\hat{J}(k), e^{1}} }{g\pqty{\hat{J}(k), e^{0}}}=-\frac{J^{r}}{J_{t}}.
\end{align}
Using the inverse formula $\operatorname{arctanh}y =\frac{1}{2}\log\pqty{\frac{1+y}{1-y}}$, we obtain
\begin{align}
\Psi(k) &=\operatorname{arctanh}\pqty{-\frac{J^{r}}{J_{t}}}=\frac{1}{2}\log\pqty{\frac{J_{t}-J^{r}}{J_{t}+J^{r}}}_{k}, \label{PsilogJ}\\
\psi(k) &=\Psi(k)-\Psi(0)=\frac{1}{2}\log\bqty{ \pqty{\frac{J_{t}-J^{r}}{J_{t}+J^{r}}}_{k} \cdot \pqty{\frac{J_{t}-J^{r}}{J_{t}+J^{r}}}^{-1}_{k=0}},
\end{align}
where the subscripts $k$ and $k=0$ indicate that the quantities $J_{t}$ and $J^{r}$ are taken with the electromagnetic coupling parameter $k$ and with $k=0$ (neutral gas), respectively.

On the other hand, note that
\begin{align}
\pqty{\frac{J_{t}-J^{r}}{J_{t}+J^{r}}}_{\text{plasma}}&=\tilde{\lambda}_{1}  \pqty{\frac{J_{t}-J^{r}}{J_{t}+J^{r}}}_{k_{1}}+\tilde{\lambda}_{2}  \pqty{\frac{J_{t}-J^{r}}{J_{t}+J^{r}}}_{k_{2}},\\
\tilde{\lambda}_{\ell}&\equiv \lambda_{\ell} \pqty{ \frac{J_{t}+J^{r}}{n_{\ell\infty}}}_{k_{\ell}} \cdot \bqty{ \sum_{\ell'} \lambda_{\ell'} \pqty{ \frac{J_{t}+J^{r}}{n_{\ell'\infty}}}_{k_{\ell'}}}^{-1},
\end{align}
from which, together with the expression for $\psi(k)$, it immediately follows that $\Bqty{\psi_{1}, \psi_{2}}_{\min} < \psi_{\text{plasma}} < \Bqty{\psi_{1}, \psi_{2}}_{\max}$.

We now analyze the asymptotic behavior of $\psi(k)$ at the horizon. As $r \to r^{+}_{H}$, $h(r) \to 0^{+}$, and we have
\begin{align}
\psi(k) &=\frac{1}{2}\log\bqty{\frac{A(0)B(k)}{A(k)B(0)}},\\
 A(k)&\equiv \int^{\infty}_{1}f_{\infty}(E)L^{2}_{c}(E, k) \dd{E},\notag\\
 B(k)&\equiv \int^{\infty}_{1}f_{\infty}(E) L^{2}_{c}(E, k) \frac{L^{2}_{c}(E, k)+2r^{2}_{H}}{(E-k/r_{H})^{2}}\dd{E}.\notag
\end{align}

If we take the simple monoenergetic distribution $f \propto \delta(E-\epsilon)$, we obtain
\begin{align}
\psi(k)=\frac{1}{2}\log \bqty{ \frac{L^{2}_{c}(\epsilon, k)+2r^{2}_{H}}{L^{2}_{c}(\epsilon, 0)+2r^{2}_{H}}\cdot \pqty{1-\frac{k}{\epsilon r_{H}}}^{-2}}.
\end{align}
Obviously, as $\epsilon \to \infty$, $\psi(k) \to 0$, hence $\Delta \psi \to 0$. In this limit the property of $L_{c}(E, k)$ discussed in Sec.~\ref{Sec2} has been used, namely the limit Eq.~\eqref{limbcLbyE}.

\begin{figure*}[htbp]
\begin{minipage}{0.48\textwidth}  
{ \centering \includegraphics[width=\linewidth]{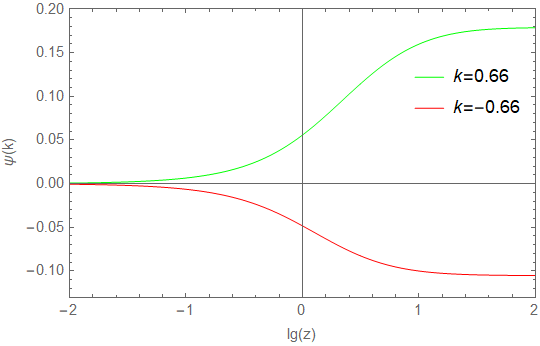} }
\caption{Relative rapidity $\psi(k)$ as a function of the relativistic coldness $z$ for the J\"uttner distribution $f_{\infty}\propto e^{-z E}$. The plot parameter is $Q=0.66$.}
\label{fig:psiJ}
\end{minipage}\hfill
\begin{minipage}{0.48\textwidth}
{ \centering  \includegraphics[width=\linewidth]{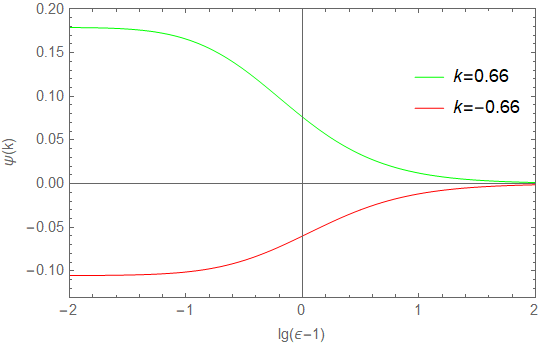} }
\caption{Relative rapidity $\psi(k)$ as a function of the energy $\epsilon$ for the monoenergetic distribution $f_{\infty}\propto \delta(E-\epsilon)$. The plot parameter is $Q=0.66$.}
\label{fig:psiepsilon}
\end{minipage}
\end{figure*}

Figs.~\ref{fig:psiJ} and~\ref{fig:psiepsilon} display the curves of $\psi(k)$ for the J\"uttner and monoenergetic distributions, respectively, with $k=\pm 0.66$. The numerical results indicate that for electromagnetic repulsion ($k=0.66$), $\psi(k)>0$; for electromagnetic attraction ($k=-0.66$), $\psi(k)<0$. This means that in the neutral-gas reference frame the electromagnetically repulsive component has motion along the $\partial_{r}$ direction, while the attractive component has motion in the opposite direction; in other words, electromagnetic repulsion suppresses the inward flow toward the black hole, while electromagnetic attraction enhances it.

On the other hand, for typical electromagnetic coupling strengths $Q=0.66$, $k=\pm 0.66$, the numerical curves show that $\frac{(\Delta \psi)^{2}}{2}$ is always a small quantity, which is a sufficient condition for the validity of the approximate formula Eq.~\eqref{nlambdann} in the main text.

\section{Eigenvector decomposition of the stress-energy tensor}\label{AppB}

We expand the stress-energy tensor $T_{ab}$ in the static observer frame:
\begin{align}
T &=T_{00}e^{0}\otimes e^{0}+ T_{11}e^{1}\otimes e^{1}+T_{01}\pqty{ e^{0}\otimes e^{1} +e^{1}\otimes e^{0}}\notag\\
&\quad +	T_{22}e^{2}\otimes e^{2}+ T_{33}e^{3}\otimes e^{3}. \label{Temuenu}
\end{align}

Let $w_{\alpha}$ ($\alpha=0,1,2,3$) be the eigenvectors of $T^{a}{}_{b}$, i.e., $T^{a}{}_{b}(w_{\alpha})^{b}=\lambda_{\alpha}(w_{\alpha})^{a}$. Because of the rotational symmetry in the $e_{2},e_{3}$ directions, the choices $w_{2}=e_{2}$ and $w_{3}=e_{3}$ automatically satisfy the eigenvector equations. In the $\operatorname{span}\{e_{0}, e_{1}\}$ subspace, the timelike and spacelike eigenvectors can be parametrized as
\begin{align}
 w_{0}& = \cosh\Theta e_{0}+\sinh\Theta e_{1}, &
 w_{1} &= \sinh\Theta e_{0}+\cosh\Theta e_{1},
\end{align}
where the rapidity $\Theta$ satisfies
\begin{align}
\cosh2\Theta &=\frac{T_{00}+T_{11}}{D}, &
\sinh2\Theta &=\frac{-2T_{01}}{D}, &
\tanh2\Theta &=\frac{-2T_{01}}{T_{00}+T_{11}},\label{tanhTheta}
\end{align}
with $D \equiv \sqrt{(T_{00}+T_{11})^{2}-4T^{2}_{01}}$. The corresponding dual frame is
\begin{align}
 w^{0}& = \cosh\Theta e^{0}-\sinh\Theta e^{1}, &
 w^{1} &= -\sinh\Theta e^{0}+\cosh\Theta e^{1},\\
 w^{2}&=e^{2}, & w^{3}&=e^{3}.
\end{align}

In this frame, $T_{ab}$ takes the diagonal form
\begin{align}
T&=\varepsilon w^{0}\otimes w^{0}+p_{\text{rad}}w^{1} \otimes w^{1}+p_{\text{tan}}\pqty{ w^{2} \otimes w^{2}+w^{3}\otimes w^{3}}, 
\end{align}
where the proper energy density, proper radial pressure, and proper tangential pressure are given by
\begin{align}
\varepsilon &=\frac{ D+(T_{00}-T_{11}) }{2}, &
p_{\text{rad}}&=\frac{ D-(T_{00}-T_{11}) }{2}, &
p_{\text{tan}} &=T_{22}. \label{equation1}
\end{align}

We now consider a two-component system. For component $\ell$ (coupling parameter $k_{\ell}$), we denote its eigenvectors and rapidity by $w^{(\ell)}_{\alpha}=w_{\alpha}(k_{\ell})$ and $\Theta_{\ell}=\Theta(k_{\ell})$; the corresponding quantities for the neutral gas ($k=0$) are denoted with an underline: $\underline{w}_{\alpha}$, $\underline{\Theta}$. The total stress-energy tensor of the plasma is $T_{ab}=\sum_{\ell}T^{(\ell)}_{ab}$.

Define the rapidity difference of component $\ell$ with respect to the neutral-gas frame $\Bqty{\underline{w}_{\alpha}, \alpha=0,1,2,3}$ as
\begin{align}
\vartheta_{\ell} \coloneqq  \Theta_{\ell}-\underline{\Theta},
\end{align}
then the dual frames of the two components can be expressed through $\underline{w}^{\alpha}$ as
\begin{align}
w^{0}_{(\ell)}&=\cosh \vartheta_{\ell}\underline{w}^{0}-\sinh \vartheta_{\ell}\underline{w}^{1}, &
w^{1}_{(\ell)}&=- \sinh\vartheta_{\ell}\underline{w}^{0}+\cosh \vartheta_{\ell}\underline{w}^{1}.
\end{align}

Expanding $T^{(\ell)}_{ab}$ in the basis $\{\underline{w}^{\alpha}\}$, we obtain
\begin{align}
T^{(\ell)}&=\pqty{\varepsilon_{\ell}\cosh^{2}\vartheta_{\ell}+p_{\ell\text{rad}}\sinh^{2} \vartheta_{\ell}}\underline{w}^{0}\otimes \underline{w}^{0}\notag\\
	&\quad +\pqty{p_{\ell\text{rad}}\cosh^{2}\vartheta_{\ell}+\varepsilon_{\ell}\sinh^{2} \vartheta_{\ell}} \underline{w}^{1} \otimes \underline{w}^{1}\notag\\
	& \quad  -(\varepsilon_{\ell}+p_{\ell\text{rad}})\cosh\vartheta_{\ell}\sinh\vartheta_{\ell}\pqty{ \underline{w}^{0}\otimes \underline{w}^{1}+\underline{w}^{1}\otimes \underline{w}^{0}}.
\end{align}
Introduce the following combinations
\begin{align}
A_{\ell}&\equiv T^{(\ell)}\pqty{\underline{w}^{0}, \underline{w}^{0}}+T^{(\ell)}\pqty{\underline{w}^{1}, \underline{w}^{1}}=\pqty{ \varepsilon_{\ell}+p_{\ell\text{rad}}}\cosh2\vartheta_{\ell},\\
B_{\ell}&\equiv T^{(\ell)}\pqty{\underline{w}^{0}, \underline{w}^{0}}-T^{(\ell)}\pqty{\underline{w}^{1}, \underline{w}^{1}}= \varepsilon_{\ell}-p_{\ell\text{rad}},\\
C_{\ell}&\equiv -2T^{(\ell)}\pqty{\underline{w}^{0}, \underline{w}^{1}}=\pqty{ \varepsilon_{\ell}+p_{\ell\text{rad}}}\sinh2\vartheta_{\ell},
\end{align}
so that $\vartheta_{\ell}$ and the intrinsic quantities of the component are given by
\begin{align}
\tanh 2\vartheta_{\ell}& =\frac{C_{\ell}}{A_{\ell}}, &
\varepsilon_{\ell} &=\frac{\sqrt{A^{2}_{\ell}-C^{2}_{\ell}}+B_{\ell}}{2}, &
p_{\ell \text{rad}}& =\frac{\sqrt{A^{2}_{\ell}-C^{2}_{\ell}}-B_{\ell}}{2}.
\end{align}

For the total stress-energy tensor $T=T^{(1)}+T^{(2)}$, its rapidity $\vartheta$ (relative to $\underline{w}_{0}$) satisfies
\begin{align}
\tanh 2\vartheta &=\frac{C_{1}+C_{2}}{A_{1}+A_{2}}\notag\\
&=\frac{ (\varepsilon_{1}+p_{1\text{rad}})\cosh2\vartheta_{1}}{\sum_{\ell}\pqty{ \varepsilon_{\ell}+p_{\ell\text{rad}}}\cosh2\vartheta_{\ell}}\tanh2\vartheta_{1}
+\frac{ (\varepsilon_{2}+p_{2\text{rad}})\cosh2\vartheta_{2}}{\sum_{\ell}\pqty{ \varepsilon_{\ell}+p_{\ell\text{rad}}}\cosh2\vartheta_{\ell}}\tanh2\vartheta_{2},
\end{align}
from which it is evident that $\min\{\vartheta_{1},\vartheta_{2}\} < \vartheta < \max\{\vartheta_{1},\vartheta_{2}\}$.

The total energy density and total radial pressure are
\begin{align}
\varepsilon & =\frac{D+(B_{1}+B_{2})}{2}, &
p_{\text{rad}} &=\frac{D-(B_{1}+B_{2})}{2},
\end{align}
where
\begin{align}
D^{2} &\equiv (A_{1}+A_{2})^{2}-(C_{1}+C_{2})^{2} \notag\\
&=\pqty{ \varepsilon_{1}+p_{1\text{rad}}+\varepsilon_{2}+p_{2\text{rad}}}^{2}
+2(\varepsilon_{1}+p_{1\text{rad}})(\varepsilon_{2}+p_{2\text{rad}})\bqty{\cosh2(\vartheta_{1}-\vartheta_{2})-1}.
\end{align}

Similar to the property of the particle current rapidity difference $\Delta\psi$, at infinity $\vartheta_{\ell}\to 0$, and for a non-cold plasma $\Delta\vartheta=\vartheta_{1}-\vartheta_{2}$ is a small quantity. Expanding $D$ to order $(\Delta\vartheta)^{2}$, we obtain the approximate formulas
\begin{align}
\varepsilon  &\simeq \pqty{\varepsilon_{1}+\varepsilon_{2}}+\frac{(\varepsilon_{1}+p_{1\text{rad}})(\varepsilon_{2}+p_{2\text{rad}})}{ \varepsilon_{1}+p_{1\text{rad}}+\varepsilon_{2}+p_{2\text{rad}}}\cdot (\Delta \vartheta)^{2}, \label{Appepsilon}\\
p_{\text{rad}}&\simeq \pqty{p_{1\text{rad}}+p_{2\text{rad}}}+\frac{(\varepsilon_{1}+p_{1\text{rad}})(\varepsilon_{2}+p_{2\text{rad}})}{ \varepsilon_{1}+p_{1\text{rad}}+\varepsilon_{2}+p_{2\text{rad}}}\cdot (\Delta \vartheta)^{2}. \label{Appprad}
\end{align}

Finally, from Eq.~\eqref{tanhTheta} we can obtain an explicit expression for the rapidity:
\begin{align}
\Theta(k) &=\frac{1}{4}\log \pqty{\frac{T_{00}+T_{11}-2T_{01}}{T_{00}+T_{11}+2T_{01}}}_{k}
          =\frac{1}{4}\log \pqty{\frac{T_{tt}+T^{rr}-2T^{r}{}_{t}}{T_{tt}+T^{rr}+2T^{r}{}_{t}}}_{k}, \label{varthetalogT}
\end{align}
from which $\vartheta(k)=\Theta(k)-\Theta(0)$ can be written as
\begin{align}
\vartheta(k)&=\frac{1}{4}\log \bqty{ \pqty{\frac{T_{00}+T_{11}-2T_{01}}{T_{00}+T_{11}+2T_{01}}}_{k} \cdot \pqty{\frac{T_{00}+T_{11}-2T_{01}}{T_{00}+T_{11}+2T_{01}}}^{-1}_{k=0}}.
\end{align}
In the horizon limit $r\to r_{H}$, using the integral expressions from the main text we find
\begin{align}
\vartheta(k)&=\frac{1}{4}\log\pqty{\frac{\mathcal{A}(0)\mathcal{B}(k)}{\mathcal{A}(k) \mathcal{B}(0)} },\\
\mathcal{A}(k)&\equiv \int^{\infty}_{1}f_{\infty}(E) (E-k/r_{H})L^{2}_{c}(E, k)\dd{E},\\
\mathcal{B}(k)&\equiv \int^{\infty}_{1} f_{\infty}(E) \frac{L^{6}_{c}+3L^{4}_{c}r^{2}_{H}+3L^{2}_{c}r^{4}_{H}}{(E-k/r_{H})^{3}}\dd{E}.
\end{align}
For a monoenergetic distribution $f \propto \delta(E-\epsilon)$, this reduces to
\begin{align}
\vartheta(k) &=\frac{1}{4}\log\bqty{ \frac{L^{4}_{c}(\epsilon, k)+3L^{2}_{c}(\epsilon, k)r^{2}_{H}+3r^{4}_{H}}{L^{4}_{c}(\epsilon, 0)+3L^{2}_{c}(\epsilon, 0)r^{2}_{H}+3r^{4}_{H}} \cdot \pqty{1- \frac{k}{\epsilon r_{H}}}^{-4}},
\end{align}
and when $\epsilon\to\infty$, $\vartheta(k)\to 0$, so that $\Delta\vartheta\to 0$.

\begin{figure*}[htbp]
\begin{minipage}{0.48\textwidth}  
{ \centering \includegraphics[width=\linewidth]{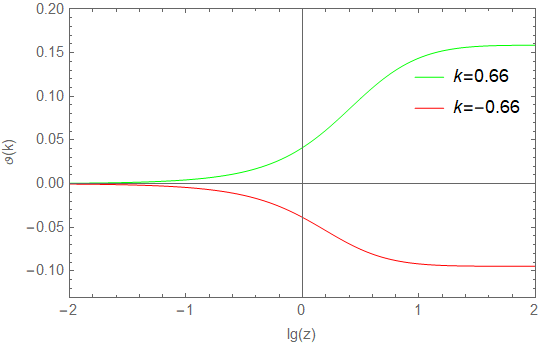}}
\caption{Relative rapidity $\vartheta(k)$ as a function of the relativistic coldness $z$ for the J\"uttner distribution $f_{\infty}\propto e^{-z E}$. The plot parameter is $Q=0.66$.}
\label{fig:varthetaJ}
\end{minipage}\hfill
\begin{minipage}{0.48\textwidth}
{ \centering  \includegraphics[width=\linewidth]{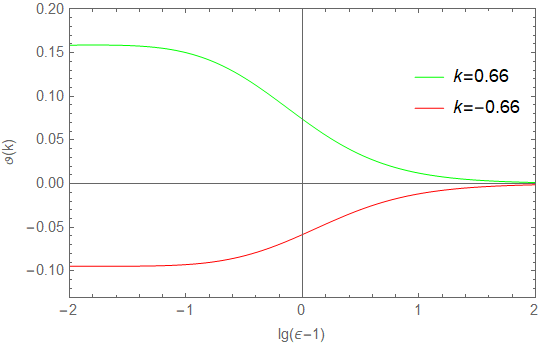} }
\caption{Relative rapidity $\vartheta(k)$ as a function of the energy $\epsilon$ for the monoenergetic distribution $f_{\infty}\propto \delta(E-\epsilon)$. The plot parameter is $Q=0.66$.}
\label{fig:varthetaepsilon}
\end{minipage}
\end{figure*}

Figs.~\ref{fig:varthetaJ} and~\ref{fig:varthetaepsilon} show the numerical curves of $\vartheta(k)$ for the J\"uttner and monoenergetic distributions (taking $k=\pm 0.66$), respectively. The results demonstrate that for conventional electromagnetic coupling strengths $(\Delta\vartheta)^{2}$ is indeed a small quantity, which is a sufficient condition  for the validity of the approximate formulas~\eqref{varepsilonsimeq} and~\eqref{pradsimeq} in the main text.

\nocite{*}
\bibliographystyle{unsrt}
\bibliography{apssamp}

\end{document}